\documentclass[%
 reprint,
superscriptaddress,
 amsmath,amssymb,
 aps,
]{revtex4-2}

\usepackage{graphicx}
\usepackage{dcolumn}
\usepackage{bm}
\usepackage[T1]{fontenc}
\usepackage{mathtools}
\usepackage{tikz}
\usepackage{soul}

\newcommand{\dbar}{d\hspace*{-0.08em}\bar{}\hspace*{0.1em}}

\newcommand*\circled[1]{\tikz[baseline=(char.base)]{\node[shape=circle,draw,inner sep=2pt] (char) {#1};}}
\newcommand{\RN}[1]{\uppercase\expandafter{\romannumeral #1\relax}}

\usepackage{etoolbox}
\appto\appendix{\counterwithin{equation}{section}}

\begin{document}

\preprint{APS/123-QED}

\title{A field theory approach to the statistical kinematic dynamo}

\author{Daria Holdenried-Chernoff}
\email{dariah@berkeley.edu}
\affiliation{Department of Earth and Planetary Science, University of California, Berkeley
}
\author{David A. King}%
\affiliation{Department of Physics and Astronomy, University of Pennsylvania, 209 South 33rd St., Philadelphia, PA, 19104.
}
\author{Bruce A. Buffett}
\affiliation{Department of Earth and Planetary Science, University of California, Berkeley
}
\date{\today}

\begin{abstract}
	Variations in the geomagnetic field occur on a vast range of time scales, from milliseconds to millions of years. The advent of satellite measurements has allowed for detailed studies of the short timescale geomagnetic field behaviour, but understanding the long timescale evolution remains challenging due to the sparsity of the paleomagnetic record.
This paper introduces a field theory framework for studying magnetic field generation as a result of stochastic fluid motions.
By constructing a stochastic kinematic dynamo model, we derive statistical properties of the magnetic field that may be compared to observations from the paleomagnetic record.
The fluid velocity is taken to act as a random forcing obeying Gaussian statistics. Using the Martin-Siggia-Rose-Janssen-de Dominicis (MSRJD) formalism, we compute the average magnetic field response function. From this we obtain an estimate for the turbulent contribution to the magnetic diffusivity, and find that it is consistent with results from mean-field dynamo theory. 
This framework presents much promise for studying the geomagnetic field in a stochastic context.
\end{abstract}
\maketitle
\onecolumngrid
\section{Introduction}
The Earth's magnetic field, also known as the geomagnetic field, is thought to be generated by the movement of conductive fluid in the outer core. Snapshots of the ancient geomagnetic field have been preserved in a variety of different rocks, known in their ensemble as the paleomagnetic record. Paleomagnetic evidence suggests that the geodynamo has been active for at least 3.5 billion years \cite{Biggin11, Borlina20}. The geomagnetic field has not remained static throughout its lifetime, varying on a wide range of timescales, from milliseconds to millions of years. 
	
Recent satellite data have enabled detailed studies of the magnetic field's variations on decadal timescales, revealing a rich variety of different waves in the core \cite{Gillet10,Aubert19,Buffett19,Chi21,Gillet22}. Studies of the magnetic field's variations on longer timescales are however more challenging, due to a dearth of observations older than a few million years. Furthermore, much of these data do not provide a complete description of the magnetic field, but rather catalogue variations in the 
dipole component of the field \cite{Ziegler11, Cals10k, Panovska15}. 

As we do not have a complete description of the geomagnetic field variations over long timescales, it is impossible to create exact models of the geomagnetic field and its history. Nonetheless, the data that we do have may be sufficient to understand the long timescale behaviour of the geodynamo in a statistical manner.
Imagine finding an exact copy of planet Earth, or re-starting the Earth's evolution from its formation with the same initial conditions. The processes governing the core flow and magnetic field's evolution are so complex that the chances of the Earth-copy having same magnetic field and velocity structure as the current Earth, at a given age, are essentially zero. Although the system's governing equations are known, an enormous number of parameters can influence its evolution. Consider for instance the fluid velocity in the outer core. 
Amongst other things, the fluid flow will be affected by local variations in temperature, composition and core-mantle boundary topography\cite{Jones11,Kuang01}. Attempting to capture all of these effects exactly is a futile task, similarly to trying to model all the forces acting on a Brownian particle (e.g. \cite{DavidsonTurbulence}). More constructively, the geodynamo, like Brownian motion, may be thought of as being stochastic in nature, so that the combined effect of the turbulent and non-linear interactions may be approximated by adding random noise to the velocity term in the governing equations. This transforms the  equations into stochastic differential equations (SDEs).

Two complimentary approaches have so far been used to study the stochastic dynamo problem. The first approach \cite{Buffett13} assumes that the axial dipole moment (ADM) satisfies a  SDE of the form
\begin{align}
\label{eq:SDE}
    \frac{d X}{dt} = v(X) + g(X)\zeta(t),
\end{align}
where $X$ is the ADM, whose time evolution is controlled by two terms. The first, $v(X)$, is known as the drift term, and controls the system's slow, deterministic adjustment towards equilibrium states. The second, $g(X)\zeta(t)$, is the noise term, and governs the amplitude of rapid, random fluctuations.  Various studies \cite{Guyodo99, Valet05,Cals10k,Ziegler11} have compiled paleomagnetic observations that may be used to extract statistical properties of the magnetic field, such as the timescale of its correlations or the mean axial dipole moment \cite{Buffett13, MeduriWicht16, BuffettPuranam17,MorzfeldBuffett19,DavisBuffett22}.
The drift and noise term in (\ref{eq:SDE}) can be directly fit from these observations. 
This is a useful way of extracting statistical information from the paleomagnetic record, but it means that the physical interpretation of the results somewhat delicate, as the drift and noise terms are not directly derived from an underlying model of the governing physics. These terms can however be linked to physical models by building stochastic differential equations starting from the governing equations, as done for example in \cite{Hoyng01, ScullardBuffett18}. The stochastic ADM models are thus very useful in bridging the comparisons between the physical stochastic models and the paleomagnetic observations.

This paper takes the second approach, building a physical model for a stochastically forced magnetic field as a starting point from which to compare to paleomagnetic observations.
We ask a simple question: what are the magnetic field statistics we expect to observe, assuming that the fluid velocity satisfies certain statistical properties?

For a given velocity field $\mathbf{u}(\mathbf{x},t)$, the evolution of the magnetic field $\mathbf{B}(\mathbf{x},t)$  is described by the induction equation
\begin{align}
    \label{eq:IndEq}
    \frac{\partial \mathbf{B}}{\partial t} = \nabla \times \big( \mathbf{u} \times \mathbf{B} \big) + \eta \nabla^2 \mathbf{B},
\end{align}
where $\eta$ is the molecular magnetic diffusivity.
In the true physical system, the velocity changes over time, being itself affected by the magnetic field that it generates. The velocity's variations are described by the Navier-Stokes equation, which is coupled to the induction equation through the Lorentz force. This non-linear coupling makes the system of equations challenging to solve. For this reason, we will restrict our attention to the ``kinematic dynamo problem'',  where the velocity is taken to be prescribed and the magnetic field's back reaction is neglected\cite{Moffatt78}.
However, in contrast to traditional kinematic dynamo studies, we will not specify a particular functional form for the velocity, but rather treat it as a random forcing with well-defined statistical properties. 

The problem that we have set out is very similar to a study by Kazantsev\cite{Kazantsev}, who developed a model for magnetic field generation by an isotropically turbulent fluid with infinitesimally short time correlations. This is a problem of fundamental interest, that can be extended to include further complexities, such as allowing for general velocity time-correlations \cite{Vainshtein80}, or compressibility of the fluid \cite{Schekochihin02}, which is of great relevance in astrophysical contexts.
Our methodologies differ somewhat from those used in previous work; we aim to cast Kazantsev's study in a more modern framework, which may be used to extend the results to the geodynamo.
Making use of the ``Martin-Siggia-Rose-Janssen-De Dominicis" (MSRJD) formalism \cite{MSR73, Janssen76, DeDominicis75}, we shall link the velocity's statistical properties (e.g. its correlation function) to those of the magnetic field. 
Averaging this probability over an ensemble of different realisations of the velocity then allows us to compute information about the expectation value of the magnetic field observables.
In essence, the problem will be formulated as an interacting field theory, allowing us to import a vast literature of results and analytical machinery from other disciplines. 
Beyond this boon, we hope that by framing the geophysical problem in this manner, we may introduce it to a broader community where these techniques are commonly used.

A problem of particular interest to us, as it has direct applications to data from the paleomagnetic record, is understanding the effect of turbulence on the outer core's magnetic diffusivity.
A number of studies \cite{Hoyng02, BuffettPuranam17, MorzfeldBuffett19, SadhasivanConstable22, BuffettAveryDavis22} have shown that, using current estimates for the core's molecular magnetic diffusivity, the time correlations of the magnetic field decay more rapidly than expected for a purely dipolar decay. Davis and Buffett \cite{DavisBuffett22} concluded that this rapid decay is not due to a contribution from higher order decay modes, 
implying that the higher magnetic diffusivity observed in the paleomagnetic record could be attributed to an enhancement of the magnetic diffusivity due to turbulent fluid effects\cite{BuffettHoldenried22}. In this case, the ability to physically link statistical properties of the magnetic and velocity fields under different assumptions allows us to infer core-properties that are not directly observable, such as the ensemble averaged root-mean squared velocity in the bulk of the outer core. This is particularly significant, since traditional geomagnetic inversion methods only allow us to probe the velocity at the core's surface. The value of developing different approaches to the geodynamo problem lies in establishing new physical connections that allow us to link observables to properties that can only be inferred indirectly.
 
Section 2 begins by setting out the basic problem of a magnetic field driven by random velocity fluctuations, and outlining the assumptions that are used to define the velocity's probability distribution. The MSRJD procedure for computing the magnetic field's statistics is outlined in section 3. As a first step, we derive an expression for the expectation value of a general magnetic field observable. 
In section 4 we compute the average response (Green function) to first order in the interaction, which enables us to make a new estimate for the effective magnetic diffusivity (section 5), taking into account both molecular and turbulent diffusivity contributions.
The geophysical implications of our results are discussed in section 6, where we show that our results agree exactly with mean-field theory predictions. An extension of the average response and turbulent diffusivity calculation to all orders is treated in section 7. Our conclusions are finally presented in section 8.

\section{Basic problem}

We wish to study the statistics of a magnetic field $\mathbf{B}(\mathbf{x},t)$, under the influence of isotropic turbulence. The evolution of the magnetic field is governed by the induction equation (\ref{eq:IndEq}),
subject to the usual requirement that 
\begin{align}
\label{eq:DivB}
    \nabla \cdot \mathbf{B}=0.
\end{align}
Turbulent effects enter the equation through the velocity, $\mathbf{u}(\mathbf{x},t)$, which is not prescribed to a single functional form, but rather acts as a random forcing with well-defined Gaussian statistics. The velocity's probability distribution is taken to be
 \begin{align}
 \label{eq:ProbDistUPhys}
    p[\mathbf{u}] = N \exp\bigg[-\frac{1}{2} \int d \mathbf{x}\ d \mathbf{x}'\ dt\ dt'\ u_j (\mathbf{x},t) C^{-1}_{jk} (\mathbf{x}-\mathbf{x}', t-t') u_k (\mathbf{x}',t') \bigg],
\end{align}
where $N$ is the normalisation. This distribution has vanishing mean
\begin{align}
    \langle \mathbf{u} (\mathbf{x},t) \rangle_u = \mathbf{0},
\end{align}
and correlations
\begin{align}
     \langle u_j(\mathbf{x},t) u_k(\mathbf{x}', t')\rangle_u = C_{jk}(\mathbf{x}-\mathbf{x}', t-t').
\end{align}
The angle brackets $\langle \ \dots \rangle_u$ denote averaging over all possible realisations of the velocity field. The velocity's correlation function in space and time is $C_{jk}(\mathbf{x}-\mathbf{x}',t-t')$. Note that we have chosen the correlation to have both space and time translation invariance, so that we are considering an isotropic medium.
The incompressibility condition $\nabla \cdot \mathbf{u}=0$ is imposed by requiring $\partial_j C_{jk} = \partial_k C_{jk} = 0$.
It is of course possible to include a mean fluid flow, but we shall restrict our attention to the zero-mean case.
Throughout this paper, the convention will be to denote the arguments of functions using parentheses (e.g. $f(\mathbf{x})$), while the arguments of functionals will be denoted with square brackets, as in $F[f(\mathbf{x})]$.
In this study we do not consider the spherical geometry relevant to planetary bodies, but rather assume that fields are distributed over all space.
The next section outlines the MSRJD procedure for computing magnetic field observables given these velocity statistics.

\section{MSRJD procedure} \label{sec:MSRJD}

Our aim is to compute observables of the magnetic field given eq. (\ref{eq:IndEq}) and the statistics of $\mathbf{u}$, subject to the constraint (\ref{eq:DivB}). 
One way of ensuring that the latter constraint is satisfied is to express $\mathbf{B}$ in terms of the magnetic vector potential $\mathbf{A}(\mathbf{x},t)$, where $\mathbf{B} = \nabla \times \mathbf{A}$. Written in terms of $\mathbf{A}$, the induction equation becomes
\begin{align}
\label{eq:CurlAEq}
    \nabla \times \left[ \frac{\partial \mathbf{A}}{\partial t} - \mathbf{u}\times(\nabla\times\mathbf{A}) - \eta \nabla^2 \mathbf{A} \right] = \mathbf{0}.
\end{align}
The term in brackets must be equal to the gradient of an arbitrary scalar, $\chi$, so the evolution of the vector potential is described by
\begin{align}
\label{eq:AEq}
         \frac{\partial \mathbf{A}}{\partial t} =  \mathbf{u}\times(\nabla\times\mathbf{A}) + \eta \nabla^2 \mathbf{A} +\nabla\chi.
\end{align}
Although we are interested in the statistics of the magnetic field, it is simpler to proceed with the problem in terms of the magnetic vector potential. Complete information on the latter will allow us to compute the former. 

Let us begin by considering how to compute the expectation value of an observable $\langle O[\mathbf{A}] \rangle_u$. This will in general be a functional of $\mathbf{A}$, most commonly consisting of a combination of products of $\mathbf{A}$ and/or its derivatives at various points in space and time. For example, one of the most common observables to consider is the auto-correlation
\begin{align*}
    \langle \mathbf{A}_i(\mathbf{x}_1,t_1) \mathbf{A}_j(\mathbf{x}_2,t_2) \rangle_u,
\end{align*}
while an example of a more complex observable is the expectation value of the magnetic field's helicity
\begin{align*}
    \langle \big(\nabla \times \mathbf{A}(\mathbf{x},t)\big)\cdot \big(\nabla \times \nabla \times \mathbf{A}(\mathbf{x},t) \big) \rangle_u.
\end{align*}
Evaluating these averages requires knowledge of the probability of finding a particular function $\mathbf{A}$ at all points in space and time. This can be done by using functional methods borrowed from field theory. We will employ these methods without providing their proofs (which can be found in \cite{PeskinSchroder,AltlandSimons,QFTGiftedAmateur}), though we will physically motivate their use.

It is easiest to begin by considering the problem under a discrete representation. Suppose that we split time into a discrete set of points $t_i$ labelled by an index ``$i$''. The points $t_i$ and $t_{i-1}$ are separated by an interval $\Delta t$, which vanishes in the continuous limit. 
Denoting the magnetic vector potential at each point $\mathbf{A}_i\equiv \mathbf{A}(\mathbf{x},t_i)$, we may write the expectation value of an observable as
\begin{align}
\label{eq:OADiscrete}
       \left\langle O\left[\{\mathbf{A}_i\}\right] \right\rangle = \frac{\displaystyle\int { \displaystyle\prod_i} d\mathbf{A}_i \ O[\{\mathbf{A}_i\}] \langle p[\{\mathbf{A}_i\}\lvert\{\mathbf{u}\}] \rangle_u}{ \displaystyle\int {\displaystyle \prod_i} d\mathbf{A}_i \ \langle p[\{\mathbf{A}_i\}\lvert\{\mathbf{u}_i\}] \rangle_u},
\end{align}
where $\{\mathbf{A}_i\}$ and $\{\mathbf{u}_i\}$ denote the set of all $\mathbf{A}_i$ and $\mathbf{u}_i$ respectively. The probability of obtaining the set of all possible realisations of the vector potential, given all possible velocity realisations, is written $\langle p[\{\mathbf{A}_i\}\lvert\{\mathbf{u}_i\}] \rangle_u$. The denominator is simply the probability distribution's normalisation.
In the continuous limit, (\ref{eq:OADiscrete}) becomes
\begin{align}
\label{eq:OA3}
       \langle O[\mathbf{A}] \rangle = 
       \frac{\int \mathcal{D}\mathbf{A}  \ O[\mathbf{A}] \langle p[\mathbf{A}\lvert\mathbf{u}] \rangle_u}{\int \mathcal{D}\mathbf{A}  \ \langle p[\mathbf{A}\lvert\mathbf{u}] \rangle_u},
\end{align}
where $\int \mathcal{D}\mathbf{A}$
denotes a functional integral over all possible $\mathbf{A}(\mathbf{x},t)$ and $p[\mathbf{A}\lvert\mathbf{u}]$ is the functional probability distribution of $\mathbf{A}$ given $\mathbf{u}$. 
For the moment, let us return to the discrete representation of eq. (\ref{eq:OADiscrete}). The average of the probability distribution is
\begin{align}
\label{eq:<P(Ai|ui)>}
    \langle p\left[\{\mathbf{A}_i\}\lvert\{\mathbf{u}_i\}\right] \rangle_u = \frac{\displaystyle\int \displaystyle\prod_i d\mathbf{u}_i p\left[\{\mathbf{u}_i\}\right] p\left[\{\mathbf{A}_i\}\lvert\{\mathbf{u}_i\}\right]}{\displaystyle\int \displaystyle\prod_i d\mathbf{u}_i p\left[\{\mathbf{u}_i\}\right]}.
\end{align}
For simplicity, the probability distribution for $\mathbf{u}_i$ is taken to be Gaussian and is given in eq. (\ref{eq:ProbDistUPhys}).
The probability $p\left[\{\mathbf{A}_i\}\lvert\{\mathbf{u}_i\}\right]$ requires that for any given $\mathbf{u}_i$, $\mathbf{A}_i$ must both solve the induction equation (\ref{eq:IndEq}) and satisfy the appropriate boundary conditions at each point in space and time. Denoting the $\mathbf{A}_i$ that solves  (\ref{eq:IndEq}) as $\mathbf{A}_i^*(\{\mathbf{u}_i\})$, we may write
\begin{align}
\label{eq:P(Ai|ui)}
    p\left[\{\mathbf{A}_i\}\lvert\{\mathbf{u}_i\}\right] = \prod_i \delta\left[\mathbf{A}_i - \mathbf{A}_i^*(\{\mathbf{u}_i\}) \right],
\end{align}
that is the probability of observing $\mathbf{A}_i$ for a given realisation of $\mathbf{u}_i$ is a delta-function on solutions to eq. (\ref{eq:AEq}).
In continuous notation, this is simply
\begin{align}
\label{eq:P[A|U]}
    p[\mathbf{A}\lvert\mathbf{u}] = \delta\big[\mathbf{A} - \mathbf{A}^*(\mathbf{u})\big].
\end{align}
Rather than solving for $\mathbf{A}^*(\mathbf{u})$, it is more convenient to transform to a new variable, $\mathbf{X}$, that represents eq. (\ref{eq:IndEq}). Using a discrete representation, we have
\begin{align}
\begin{split}
    \mathbf{X}_i = \frac{1}{\Delta t}\big(\mathbf{A}(\mathbf{x}, t_i) - \mathbf{A}(\mathbf{x}, t_{i-1})\big) -  \mathbf{u}(\mathbf{x}, t_{i-1})\times\nabla\times \mathbf{A}(\mathbf{x}, t_{i-1}) &\\
    - \eta \nabla^2 \mathbf{A}(\mathbf{x}, t_{i-1}) -\nabla\chi(\mathbf{x}, t_{i-1}).&
\end{split}
\end{align}
Note that $\mathbf{X}=\mathbf{0}$ when $\mathbf{A}=\mathbf{A}^*(\mathbf{u})$. Making use of the standard property of delta-functions, 
eq. (\ref{eq:P(Ai|ui)}) becomes
\begin{align}
    p\left[\{\mathbf{A}_i\}\lvert\{\mathbf{u}_i\}\right] = \left\lvert \frac{\partial \mathbf{X}_j}{\partial \mathbf{A}_k}\right \lvert \prod_i \delta [\mathbf{X}_i].
\end{align}
In continuous notation, we have
 \begin{align}
     p[\mathbf{A}\lvert\mathbf{u}] =  \bigg\lvert \frac{\delta \mathbf{X}}{\delta \mathbf{A}} \bigg\lvert\delta[\mathbf{X}].
 \end{align}
The Jacobian matrix $\partial \mathbf{X}_j/\partial \mathbf{A}_k$ is triangular, so its determinant equals the product of its diagonal elements. In this case, it is simple to show that $\big| \partial \mathbf{X}_j/\partial \mathbf{A}_k \big| \propto 1/(\Delta t)^N$. Since the determinant does not depend on $\mathbf{A}_i$ or $\mathbf{u}_i$, only on the choice of discretization, any constant values will cancel out with the probability distribution's normalisation once we take the averages required for (\ref{eq:OADiscrete}).
To make progress, we may write the delta function in terms of its Fourier transform. In one dimension we have
\begin{align}
    \delta (x-a) = \int dk e^{ik(x-a)},
\end{align}
which becomes 
\begin{align}
    \prod_i \delta(\mathbf{X}_i) = \int \prod_i d\tilde{\mathbf{A}}_i \exp\left[i\sum_i  \tilde{\mathbf{A}}_i \cdot \mathbf{X}_i\right]
\end{align}
in the discrete case.
Passing back to the continuous representation, we may write the probability distribution as
\begin{align}
\begin{split}
\label{eq:p[A]}
    p[\mathbf{A}\lvert\mathbf{u}] &\propto \delta \bigg[ \frac{\partial \mathbf{A}}{\partial t} - \mathbf{u}\times(\nabla\times\mathbf{A}) - \eta \nabla^2 \mathbf{A}- \nabla \chi \bigg]\\
    & \propto\int \mathcal{D} \tilde{\mathbf{A}}\exp\bigg[i\int d\mathbf{x} dt \tilde{\mathbf{A}} \cdot \bigg( \frac{\partial \mathbf{A}}{\partial t} - \mathbf{u}\times(\nabla\times\mathbf{A}) - \eta \nabla^2 \mathbf{A}- \nabla \chi \bigg) \bigg]
\end{split}
\end{align}
where the normalisation has not explicitly been written, and $\tilde{\mathbf{A}}(\mathbf{x},t)$ is the ``conjugate field'' to $\mathbf{A}(\mathbf{x},t)$ \cite{AltlandSimons}. The exponent's argument is integrated over time and space. This can be understood by considering the discrete representation of the field; at every spatio-temporal point $(\mathbf{x},t)$, each realisation of $\mathbf{A}(\mathbf{x},t)$ has an associated $\tilde{\mathbf{A}}(\mathbf{x},t)$. Summing over these realisations produces the integrals given in (\ref{eq:p[A]}). 
To obtain the average of $p[\mathbf{A}\lvert \mathbf{u}]$ over all realisations of $\mathbf{u}$ we take the continuous limit of (\ref{eq:<P(Ai|ui)>}), 
\begin{align}
\label{eq:<P>_u}
    \left\langle p[\mathbf{A}\lvert \mathbf{u}]\right\rangle_u = \frac{\displaystyle\int \mathcal{D}\mathbf{u} \ p\left[\mathbf{u}\right] p\left[\mathbf{A}\lvert \mathbf{u}\right]}{\displaystyle\int \mathcal{D}\mathbf{u} \ p\left[\mathbf{u}\right]},
\end{align}
obtaining a path integral over $\mathbf{u}$.
Combining (\ref{eq:ProbDistUPhys}), (\ref{eq:p[A]}) and (\ref{eq:<P>_u}) we find
\begin{align}
\label{eq:OA1}
\begin{split}
    \langle O[\mathbf{A}] \rangle_u = \mathcal{N} \int \mathcal{D}\mathbf{A} \mathcal{D}\tilde{\mathbf{A}}\mathcal{D}\mathbf{u}\ O[\mathbf{A}] \exp\bigg[-\frac{1}{2} \int d \mathbf{x}\ d \mathbf{x}'\ dt \ dt'\ u_k C^{-1}_{kj} u_j&\\
     + i\int d\mathbf{x} dt \tilde{A}_j \bigg( \frac{\partial {A}_j}{\partial t} -  u_k \varepsilon_{klj} (\nabla\times \mathbf{A})_l& - \eta \nabla^2 {A}_j- \nabla_j \chi \bigg)\bigg],
\end{split}
\end{align}
where the fields' arguments have been kept implicit for notational clarity, and $\mathcal{N}$ is the normalisation constant. The summation convention is used throughout, so that repeated indices are summed over. Given the form of $p[\mathbf{u}]$, we may take the functional integral over the velocity by performing a Gaussian integral. We make use of the identity (see chapter 3 of \cite{AltlandSimons})
\begin{align}
\begin{split}
     \int \mathcal{D}\mathbf{u}\  \exp\bigg[-\frac{1}{2} \int d\mathbf{x}  d\mathbf{x}'  dt dt' \mathbf{u}(\mathbf{x},t) \cdot \mathbf{C}^{-1}(\mathbf{x},\mathbf{x}',t,t') \cdot \mathbf{u}(\mathbf{x}',t') - \int d\mathbf{x}  dt \ \mathbf{h}(\mathbf{x},t)\cdot\mathbf{u}(\mathbf{x},t) \bigg]& \\ 
    = \det(2\pi \mathbf{C})^{1/2} \exp\bigg[\frac{1}{2} \int \int d\mathbf{x} \ d\mathbf{x}' \ dt \ dt' \ \mathbf{h}(\mathbf{x},t)\cdot \mathbf{C}(\mathbf{x},\mathbf{x}',t,t')\cdot \mathbf{h}(\mathbf{x}',t')\bigg]&,
\end{split}
\end{align}
to write (\ref{eq:OA1}) as
\begin{align}
\label{eq:OASuffix}
     \langle O[\mathbf{A}] \rangle =  \mathcal{N}&\int \mathcal{D} \mathbf{A} \mathcal{D} \tilde{\mathbf{A}} \ O[\mathbf{A}] \exp\bigg[ - S_0[\mathbf{A},\tilde{\mathbf{A}}] -  S_{\text{int}}[\mathbf{A},\tilde{\mathbf{A}},\mathbf{C}]\bigg].
\end{align}
Here we have introduced the action $S[\mathbf{A},\tilde{\mathbf{A}}] = S_0[\mathbf{A},\tilde{\mathbf{A}}] + S_{\text{int}}[\mathbf{A},\tilde{\mathbf{A}},\mathbf{C}]$, where $S_0[\mathbf{A},\tilde{\mathbf{A}}]$ (the ``free action'') is the contribution we would have from free decay (i.e. $\mathbf{u}=\mathbf{0}$), while $S_{\text{int}}[\mathbf{A},\tilde{\mathbf{A}},\mathbf{C}]$ (the ``interaction action'') represents the interaction part obtained by integrating over all possible representations of $\mathbf{u}$. The zero-order and velocity-dependent interaction parts are
\begin{flalign}
\begin{split}
\label{eq:Sdefs}
    S_0 = - i \int & d\mathbf{x} \ dt \ \tilde{\mathbf{A}}\cdot\big(\partial_t\mathbf{A}-\eta \nabla^2 \mathbf{A}\big)=- i \int d\mathbf{x} \ dt \ \tilde{\mathbf{A}}  \cdot G_0^{-1} \cdot \mathbf{A}, \text{ and }\\
    S_{\text{int}} = \frac{1}{2}\int & d \mathbf{x}\ d \mathbf{x}'\ dt\ dt' \\
    &\quad \left[\varepsilon_{jab} (\nabla \times \mathbf{A})_a(\mathbf{x},t) \tilde{A}_b(\mathbf{x},t)) C_{jk}(\mathbf{x},\mathbf{x}',t,t')(\varepsilon_{kcd} (\nabla \times \mathbf{A})_c (\mathbf{x}',t') \tilde{A}_d(\mathbf{x}',t')\right]
\end{split}
\end{flalign}
 respectively. Note that we have written the purely diffusive operator in terms of the inverse of its Green function, $G_0(\mathbf{x},t)$.
For our purposes, it is more convenient to Fourier transform (\ref{eq:Sdefs}) in both space and time. We relegate the transformation procedure to Appendix \ref{app:ToFourier}, and note that the velocity correlations may be written in Fourier space as
 \begin{align}
 \label{eq:UCorrFourier}
     \langle u_j(\mathbf{q},\omega) u_k(\mathbf{q}', \omega')\rangle  = (2\pi)^4  \delta(\mathbf{q}+\mathbf{q}')\delta(\omega+\omega') C_{jk}(\mathbf{q},\omega)  ,
 \end{align}
 where $\mathbf{q}$ and $\mathbf{q}'$ are wavevectors. Due to incompressibility, we may write
 \begin{align}
  \label{eq:SigmaDef}
    C_{jk}(\mathbf{q},\omega) =\gamma(\mathbf{q},\omega)\sigma_{jk}(\mathbf{q}), && \text{where} &&
     \sigma_{jk}(\mathbf{q}) = \delta_{jk} - \frac{1}{q^2} q_j q_k.
 \end{align}
This function guarantees the velocity's incompressibility, as $q_j \sigma_{jk} = q_k\sigma_{jk} = 0$. Note that the velocity correlations are symmetric upon swapping the indices $i$ and $j$.
We may relate $C_{jk}$ and $\gamma$ in real space by
\begin{align}
    C_{jk}(\mathbf{x},t) = \delta_{jk} \gamma(\mathbf{x},t)+\int d\mathbf{y} \frac{1}{4\pi \lvert \mathbf{x}-\mathbf{y}\lvert}\partial_j\partial_k \gamma(\mathbf{y},t).
\end{align}
Following the procedure outlined in Appendix \ref{app:ToFourier}, in essence an application of the convolution theorem, the Fourier transform of the interaction action may be written
\begin{align}
\begin{split}
\label{eq:InteractionAction}
    S_{\text{int}} =-\frac{1}{2} \int \dbar \mathbf{p} \dbar \mathbf{p}' \dbar \mathbf{q} \dbar \omega \dbar \omega' \dbar \Omega & \ A_\alpha (\mathbf{p},\omega) \tilde{A}_\beta (-\mathbf{p}-\mathbf{q},-\omega-\Omega) \\
    & M_{\alpha\beta\gamma\delta}^{kl}(\mathbf{q},\Omega) p_k p_l' A_\gamma (\mathbf{p}',\omega')\tilde{A}_\delta (\mathbf{q}-\mathbf{p}',\Omega - \omega'),
\end{split}
\end{align}
where $\mathbf{p}$, $\mathbf{p}'$ and $\mathbf{q}$ are newly defined wavevectors, $\omega,\omega'$ and $\Omega$ are conjugate frequencies, and the bar on the integration measure represents a division by $2\pi$ for each dimension, i.e. $\dbar \mathbf{p}= \displaystyle\frac{d\mathbf{p}}{(2\pi)^3}$ and $\dbar \omega = \displaystyle\frac{d\omega}{2\pi}$. The matrix $M_{\alpha\beta\gamma\delta}^{kl}(\mathbf{q},\Omega)$ is defined as
\begin{align}
    M_{\alpha\beta\gamma\delta}^{kl}(\mathbf{q},\Omega) = \big( \delta_{k \beta} \delta_{l \delta} \sigma_{\alpha \gamma}(\mathbf{q}) - \delta_{\alpha \beta} \delta_{l\delta} \sigma_{k\gamma}(\mathbf{q})-\delta_{k\beta}\delta_{\gamma \delta} \sigma_{\alpha l} (\mathbf{q}) + \delta_{\alpha \beta} \delta_{\gamma \delta} \sigma_{k l} (\mathbf{q})  \big) \gamma(\mathbf{q},\Omega).
\end{align}
Comparing (\ref{eq:OASuffix}) to 
\begin{align}
   \langle O[\mathbf{A}] \rangle = \int \mathcal{D}\mathbf{A} \mathcal{D}\tilde{\mathbf{A}} \ O[\mathbf{A}] p[\mathbf{A},\tilde{\mathbf{A}}],
\end{align}
we see that the newly defined probability distribution for $\mathbf{A}$ and $\tilde{\mathbf{A}}$ is 
\begin{align}
    p[\mathbf{A},\tilde{\mathbf{A}}] = \mathcal{N}\exp\big[-S_0[\mathbf{A},\tilde{\mathbf{A}}]-S_{\text{int}}[\mathbf{A},\tilde{\mathbf{A}},\mathbf{C}]\big].
\end{align}
This quantity is very useful, as it will allow us to compute observables of $\mathbf{A}$ that can directly be related to observables of the magnetic field. For example, the second moment of the magnetic field can be written in Fourier space as
\begin{equation}
\label{eq:ObsB2}
    \langle B_{a} (\mathbf{q}_1,\omega_1) B_{d}(\mathbf{q}_2,\omega_2)\rangle =- (\varepsilon_{{a}{b}{c}} q_1^{b} )(\varepsilon_{d e f } q_2^{e} ) \langle A_{c} (\mathbf{q}_1,\omega_1) A_{f} (\mathbf{q}_2,\omega_2)\rangle,
\end{equation}
while a more general observable is
\begin{equation}
\label{eq:ObsBFromObsA}
\begin{split}
    \langle B_{a_1} (\mathbf{q}_1,\omega_1) \dots B_{a_n}(\mathbf{q}_n,\omega_n)\rangle = i^n (\varepsilon_{{a_1}{b_1}{c_1}} q_1^{b_1} )\dots&(\varepsilon_{a_n b_n c_n} q_n^{b_n} )\\
    &\langle A_{c_1} (\mathbf{q}_1,\omega_1)\dots A_{c_n} (\mathbf{q}_n,\omega_n)\rangle.
\end{split}
\end{equation}
\subsection{A note on gauge invariance}
At this point, it is worthwhile pausing to consider the system's dependence on the choice of gauge. All observables of $\mathbf{B}$ are gauge invariant, so we expect the RHS of (\ref{eq:ObsBFromObsA}) to also be gauge invariant. The cross products with $\mathbf{q}$ remove $\mathbf{A}$'s gauge dependence, which means that we require $p[\mathbf{A},\tilde{\mathbf{A}}]$ to also be gauge invariant.
Given that $p[\mathbf{A},\tilde{\mathbf{A}}]$ must be gauge invariant, it is possible to transform to a situation where $\mathbf{A} = 0$, so that we get an infinite number of constant integrals that lead to a poorly defined average \cite{PeskinSchroder}. To avoid this problem, it is necessary to fix $\mathbf{A}$'s gauge. We may turn to the Faddeev-Popov\cite{FaddeevPopov} trick, commonly used when considering functional integrals of the electromagnetic field. This trick leads to the introduction of the term \cite{PeskinSchroder}
\begin{align}
\label{eq:GaugeFix}
    \exp\bigg[-\frac{i}{2g}\int d\mathbf{x} \ dt (\nabla \cdot \mathbf{A})^2\bigg]
\end{align}
to the probability (\ref{eq:p[A]}), where $g$ is a constant that determines the gauge.
The MSRJD procedure outlined above remains unaffected by the addition of the gauge fixing term in the exponent; the only difference is that $S_0$ is now defined as
\begin{align}
\label{eq:S0new}
    S_0[\mathbf{A},\Tilde{\mathbf{A}}]= -i\int d\mathbf{x} \ dt \tilde{\mathbf{A}} \cdot G_0^{-1} \cdot \mathbf{A} + \frac{i}{2g}\int d\mathbf{x} \ dt(\nabla \cdot \mathbf{A})^2,
\end{align}
where the second term in the expression for $S_0$ is the gauge fixing term (\ref{eq:GaugeFix}). It is important to ensure that the addition of this term does not affect any observables of $\mathbf{B}$. Appendix \ref{app:GaugeCheck} covers the steps required to ascertain that this remains true.
Since $\mathbf{A}$'s gauge does not influence any of the magnetic field observables we wish to calculate, we choose it such that it cancels the $\nabla\chi$ term in (\ref{eq:AEq}).

\section{Average first order response}\label{sec:FirstOrderResponse}

The MSRJD procedure has given us the means of calculating observables of the magnetic vector potential that can be linked to observables of the magnetic field. We may now compute the expected turbulent diffusivity $\eta_t$ by finding an average Green (response) function for our system. The coefficient of the diffusive term will allow us to find $\eta_t$ for this particular system. Our results may be compared directly to the work of Moffatt\cite{Moffatt78} and Kazantsev\cite{Kazantsev}, allowing us to benchmark our solution method against known results.

\subsection{Response function formalism}

We make use of the response function formalism\cite{CardyBook} to calculate the magnetic field's Green function.
Consider adding a small arbitrary forcing $\mathbf{h}$ to the equation for the magnetic vector potential, (\ref{eq:AEq})
\begin{align}
    \frac{\partial \mathbf{A}}{\partial t} =  \mathbf{u}\times(\nabla\times\mathbf{A}) + \eta \nabla^2 \mathbf{A} 
    + \mathbf{h}.
\end{align}
Note that the arbitrary potential $\nabla \chi$ has been chosen to cancel out the term guaranteeing $\mathbf{A}$'s gauge-invariance, so that it no longer appears in the equation. Solutions to this equation may be written as
\begin{align}
\label{eq:ASolnPhysSpace}
    A_\alpha(\mathbf{x},t) = \int d\mathbf{x}' dt' G_{\alpha\beta}(\mathbf{x},\mathbf{x}',t,t'){h}_\beta (\mathbf{x}',t'),
\end{align}
where $G_{\alpha\beta}(\mathbf{x},\mathbf{x}',t,t')$ is the Green function to the induction equation in physical space in the absence of the forcing $\mathbf{h}$.
Using Parseval's theorem, we may write down an equivalent expression to (\ref{eq:ASolnPhysSpace}) in Fourier space
\begin{align}
\label{eq:AGenSolnGh}
    A_\alpha(\mathbf{k},\omega) = \int d\mathbf{k}_2 \ d\omega_2 G_{\alpha\beta}(\mathbf{k},\mathbf{k}_2,\omega,\omega_2){h}_\beta (-\mathbf{k}_2,-\omega_2).
\end{align}
Note the negative sign in the arguments of $\mathbf{h}$.
As we are interested in the system's average response, we take an ensemble average of (\ref{eq:AGenSolnGh}), obtaining
\begin{align}
\label{eq:AvgAGenSolnGh}
    \left\langle A_\alpha(\mathbf{k},\omega)\right\rangle = \int d\mathbf{k}_2 \ d \omega_2 \left\langle G_{\alpha\beta}(\mathbf{k},\mathbf{k}_2,\omega,\omega_2)\right\rangle {h}_\beta (-\mathbf{k}_2,-\omega_2).
\end{align}
The forcing $\mathbf{h}(\mathbf{k},\omega)$ is independent of both $\mathbf{u}$ and $\mathbf{A}$, so the averaging does not affect it. Taking the functional derivative of the expression with respect to ${h}_\beta(-\mathbf{k}',-\omega')$, we find
\begin{align}
\begin{split}
    \left.\frac{\delta\langle{
    A}_\alpha(\mathbf{k},\omega)\rangle}{\delta {h}_\beta(-\mathbf{k}',-\omega')}\right\lvert_{\mathbf{h}=\mathbf{0}} =& \int d\mathbf{k}_2 d\omega_2 \left\langle G_{\alpha \beta} (\mathbf{k},\mathbf{k}_2,\omega,\omega_2) \right\rangle \delta(\mathbf{k}_2-\mathbf{k}')\delta(\omega_2-\omega')\\
    =& \left\langle G_{\alpha\beta}(\mathbf{k},\mathbf{k}',\omega,\omega') \right\rangle.
\end{split}
\end{align}
The LHS of this expression defines the average Green function $\langle G_{\alpha\beta}(\mathbf{k},\mathbf{k}',\omega,\omega') \rangle $. To evaluate it, we require an expression for $\langle{
    A}_\alpha(\mathbf{k},\omega)\rangle$. Repeating the steps prescribed by the MSRJD procedure, we find an expression analogous to (\ref{eq:OASuffix}), but with an additional $\mathbf{h}$-dependent factor in the exponent
\begin{align}
\begin{split}
\label{eq:RespFunctFormOA}
    \langle O[\mathbf{A}] \rangle =  \mathcal{N}\int \mathcal{D} \mathbf{A} \mathcal{D} \tilde{\mathbf{A}} \ O[\mathbf{A}] \exp\bigg[ - S_0[\mathbf{A},\tilde{\mathbf{A}}] & -  S_{\text{int}}[\mathbf{A},\tilde{\mathbf{A}},\mathbf{C}]\\
    &-i\int d{\mathbf{k}} d\omega \ \tilde{\mathbf{A}}(\mathbf{k},\omega)\cdot \mathbf{h}(-\mathbf{k},-\omega)\bigg].
\end{split}
\end{align}
Using expression (\ref{eq:RespFunctFormOA}),
the average Green function is found to be
\begin{align}
\label{eq:AvgGFourierSpace}
   \left\langle G_{\alpha\beta}(\mathbf{k},\mathbf{k}',\omega,\omega') \right\rangle= - i\langle  \mathbf{A}_{\alpha} (\mathbf{k},\omega) \tilde{\mathbf{A}}_{\beta} (\mathbf{k}',\omega')\rangle.
\end{align}
This is a very useful relation, as it allows us to compute the average Green function directly from the correlations of $\mathbf{A}$ and $\tilde{\mathbf{A}}$. 
For example, we may immediately write down the average zero-order Green function in Fourier space as
\begin{align}
\label{eq:AvgG0A}
    \langle G_{\alpha\beta}^0(\mathbf{k},\mathbf{k}',\omega,\omega') \rangle = - i\langle  \mathbf{A}_\alpha (\mathbf{k},\omega) \tilde{\mathbf{A}}_\beta (\mathbf{k}',\omega')\rangle
    = (2\pi)^4\frac{\delta_{\alpha\beta}}{i\omega + \eta k^2} \delta(\mathbf{k}+\mathbf{k}')\delta(\omega+\omega'),
\end{align}
where we have made use of (\ref{eq:AcorrsS0}).
This is the expected Green function for a purely diffusive solution.
Expression (\ref{eq:AvgGFourierSpace}) describes the vector potential's response to an impulse, rather than the magnetic field's. To understand the magnetic field's response we must relate it to $G(\mathbf{k},\mathbf{k}',\omega,\omega')$.
Assume that the Green function for the induction equation is known, and call it $\mathcal{G}(\mathbf{k},\mathbf{k}',\omega,\omega')$. The fictitious force equivalent to $\mathbf{h}$ in the induction equation is $\mathbf{f}(\mathbf{k},\omega) = i \mathbf{k} \times \mathbf{h}(\mathbf{k},\omega)$. Therefore, we may write the magnetic field solutions as:
\begin{align}
\begin{split}
    \langle \mathbf{B}_\alpha (\mathbf{k},\omega) \rangle =& \int d\mathbf{k}_2 \ d\omega_2 \langle \mathcal{G}_{\alpha \beta} (\mathbf{k},\mathbf{k}_2,\omega,\omega_2)\rangle \mathbf{f}_\beta(-\mathbf{k}_2,-\omega_2)\\
    =& -\int d\mathbf{k}_2 \ d\omega_2 \langle \mathcal{G}_{\alpha \beta}(\mathbf{k},\mathbf{k}_2,\omega,\omega_2)\rangle \big(i\mathbf{k}_2\times \mathbf{h}(-\mathbf{k}_2,-\omega_2)\big)_\beta\\
    =& -i\int d\mathbf{k}_2 \ d\omega_2 \langle \mathcal{G}_{\alpha \beta}(\mathbf{k},\mathbf{k}_2,\omega,\omega_2)\rangle\varepsilon_{\beta ij}(k_2)_i h_j (-\mathbf{k}_2,-\omega_2).
    \end{split}
\end{align}
Taking the functional derivative with respect to $\mathbf{h}_j(-\mathbf{k}',-\omega')$, we obtain
\begin{align}
\label{eq:dBdh}
    \frac{\delta \langle\mathbf{B}_\alpha (\mathbf{k},\omega)\rangle}{\delta\mathbf{h}_j(-\mathbf{k}',-\omega')} =- i \langle \mathcal{G}_{\alpha\beta}(\mathbf{k},\mathbf{k}',\omega,\omega') \rangle \varepsilon_{\beta i j} k_i'.
\end{align}
To isolate $\langle \mathcal{G}_{\alpha\beta}(\mathbf{k},\mathbf{k}',\omega,\omega') \rangle$, we operate $\mathbf{k}' \times = \varepsilon_{pnj} k_n'$ on (\ref{eq:dBdh}), and find
\begin{align}
   \langle \mathcal{G}_{\alpha p} (\mathbf{k},\mathbf{k}',\omega,\omega')\rangle =\frac{i}{(k')^2} \varepsilon_{pnj} k_n' \frac{\delta \langle B_\alpha(\mathbf{k},\omega) \rangle}{\delta h_j(-\mathbf{k}',-\omega')}.
\end{align}
Note that we have used the relation $\mathbf{k}\cdot \mathcal{G}(\mathbf{k},\mathbf{k}',\omega,\omega') = 0$, which must be true to enforce the condition $\nabla \cdot \mathbf{B}=0$.
The magnetic field is related to the vector potential by
\begin{align}
    B_\alpha(\mathbf{k},\omega) = i \mathbf{k} \times \mathbf{A}(\mathbf{k},\omega) = i \varepsilon_{\alpha l\phi} k_l A_\phi,
\end{align}
so
\begin{align}
\label{eq:GAtoGB}
    \mathcal{G}_{\alpha\beta} (\mathbf{k},\mathbf{k}',\omega,\omega')= -\frac{1}{(k')^2}\varepsilon_{\alpha l\phi}\varepsilon_{\beta n\psi} k_n' k_l G_{\psi\phi}(\mathbf{k},\mathbf{k}',\omega,\omega'),
\end{align}
where we have relabelled the free index $p\to\beta$. We may test whether the expression is correct for the zero-order case by taking the ensemble average and plugging in (\ref{eq:AvgG0A}).
We find
\begin{align}
\begin{split}
    \left\langle \mathcal{G}_{\alpha\beta} (\mathbf{k},\mathbf{k}', \omega,\omega')\right\rangle=& \frac{(2\pi)^4}{i\omega + \eta k^2}\left(\delta_{\alpha \beta}-\frac{1}{k^2}k_\alpha k_\beta\right) \delta(\mathbf{k}+\mathbf{k}')\delta(\omega+\omega')\\
    =&\frac{(2\pi)^4}{i\omega + \eta k^2}\ \sigma_{\alpha \beta}(\mathbf{k}) \ \delta(\mathbf{k}+\mathbf{k}')\delta(\omega+\omega'),
\end{split}
\end{align}
as expected for free diffusion of a divergenceless field. 
Through (\ref{eq:GAtoGB}) we have a relation between $G(\mathbf{k},\mathbf{k}', \omega,\omega')$ and $\mathcal{G}(\mathbf{k},\mathbf{k}', \omega,\omega')$. 
Let us now consider how these functions change when the interaction term is included to first order.

\subsection{Diagrams for computing interaction correlations}\label{sec:DiagsIntCorr}

Computing $\mathcal{G}(\mathbf{k},\mathbf{k}', \omega,\omega')$ involves finding the correlation of the vector potential with its conjugate field, as can be seen from eqs. (\ref{eq:AvgGFourierSpace}) and (\ref{eq:GAtoGB}). To first order, we may write
\begin{align}
\begin{split}
\label{eq:AAtilde}
   \langle A_\psi (\mathbf{k},\omega) \tilde{A}_\phi (\mathbf{k}',\omega') \rangle = \mathcal{N}\int \mathcal{D} \mathbf{A} \mathcal{D}\tilde{\mathbf{A}} \ A_\psi (\mathbf{k},\omega) \tilde{A}_\phi (\mathbf{k}',\omega') & e^{-S_0[\mathbf{A},\tilde{\mathbf{A}}]}\\
   &\left[1- S_{\text{int}}[\mathbf{A},\tilde{\mathbf{A}},\mathbf{C}] \right]+ O(S_{\text{int}}^2),
\end{split}
\end{align}
where the interaction action is given by (\ref{eq:InteractionAction}). By taking the expansion to first order, we are assuming that the magnitude of the velocity correlation function $\gamma(\mathbf{q},\omega)$ is small. 
Computing (\ref{eq:AAtilde}) to first order requires the average
\begin{align}
\label{eq:FirstOrdAvg}
    \langle A_\psi (\mathbf{k},\omega) \tilde{A}_\phi (\mathbf{k'},\omega') A_\alpha (\mathbf{p},\omega_1) \tilde{A}_\beta (-\mathbf{p}-\mathbf{q},-\omega_1-\Omega) A_\gamma (\mathbf{p}',\omega_2)\tilde{\mathbf{A}}_\delta (\mathbf{q}-\mathbf{p}',\Omega-\omega_2) \rangle.
\end{align}
As this average is Gaussian, we may employ Wick's theorem\cite{AltlandSimons} to rewrite it in terms of a sum of all possible pairwise contractions of the form
$\langle \tilde{A}_i(\mathbf{k},\omega) A_j(\mathbf{k}',\omega') \rangle.$
There are $^6C_2=15$ different combinations, some of which will vanish. The easiest way to write down these different combinations and determine all non-zero contributions is by using a diagrammatic approach pioneered by Feynman \cite{PeskinSchroder}. Each Feynman diagram represents an integral contributing to the average (\ref{eq:AAtilde}). 

It is helpful to introduce some commonly used naming conventions\cite{QFTGiftedAmateur}. The diagrams' lines can be separated into two types, \textit{external} and \textit{internal}. External lines have a free end that is not connected to anything. The free ends connect the diagram to the world external to the physical process under study. We represent this link by connecting the free end of external lines to empty circles. Vertices are the points where two or more lines join together. Internal lines are those joining two vertices together. Diagrams without external lines are known as vacuum diagrams. 

\begin{figure}
    \centering
    \includegraphics[width=\textwidth]{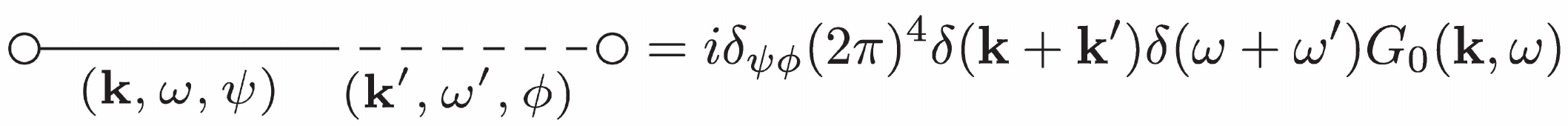}
    \caption{This diagram represents the zero order propagator, $\langle \mathbf{A}_\psi(\mathbf{k},\omega) \tilde{\mathbf{A}}_\phi(\mathbf{k}',\omega')\rangle_0$, defined in (\ref{eq:AvgG0A}). $\mathbf{A}$ is denoted by the solid line, while $\tilde{\mathbf{A}}$ is denoted by the dashed line.}
    \label{fig:ZeroDiag}
\end{figure}

Having laid out the basic terminology, we must establish a convention for our specific case. The rules for the diagrams describing our system are summarised below.
\begin{itemize}
    \item $\mathbf{A}_\psi(\mathbf{k},\omega)$ is represented by a solid line, while $\tilde{\mathbf{A}}_\phi(\mathbf{k}',\omega')$ is represented by a dashed line. Each line is labelled by its argument and the vector's index, in the form (\textit{arg}, \textit{index}).

    \item Connected lines contribute a factor of 
    \begin{align}
        (2\pi)^4 \delta(\mathbf{k}+\mathbf{k}') \delta(\omega+\omega') \ G_0(\mathbf{k},\omega),
    \end{align}
    with all $\mathbf{k}$ and $\omega$ from internal lines integrated over.

    \item Given that the auto-correlations of $\mathbf{A}$ vanish (see (\ref{eq:AcorrsS0})), the only connections that ensure a non-zero contribution are those between a solid and dashed line. 
    
    \item Vertices are denoted by a filled circle. They contribute a factor of $M_{\alpha\beta\gamma\delta}^{kl}(\mathbf{q},\Omega)p_k p_l'$ to the integral, and must be connected to four lines representing $A_\alpha(\mathbf{p},\omega_1)$, $\tilde{A}_\beta(-\mathbf{p}-\mathbf{q},-\omega_1-\Omega),A_\gamma(\mathbf{p}',\omega_2)$ and $\tilde{A}_\delta(\mathbf{q}-\mathbf{p}',\Omega-\omega_2)$. The number of vertices gives the order to which $\gamma$ appears.
\end{itemize}
The zero-order contribution to the average (\ref{eq:AAtilde}) is drawn diagrammatically in Figure \ref{fig:ZeroDiag}. The beginning and end of the diagram are denoted by two circles, connected to lines representing $A_\psi(\mathbf{k},\omega)$ and $\tilde{A}_\phi(\mathbf{k}',\omega')$ respectively. The connection between the two lines represents the average $\langle A_\psi (\mathbf{k},\omega) \tilde{A}_\phi (\mathbf{k}',\omega') \rangle_0$.

\begin{figure}
    \centering
    \includegraphics[width=0.75\textwidth]{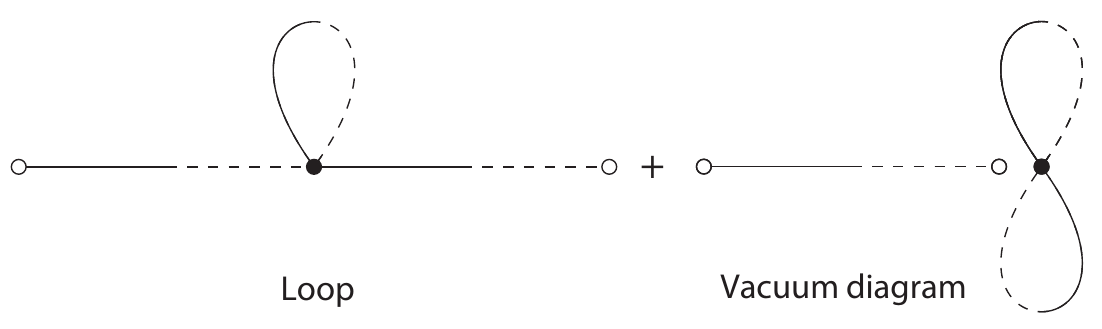}
    \caption{First order contributions to the average $\langle \mathbf{A}_\psi(\mathbf{k},\omega) \tilde{\mathbf{A}}_\phi(\mathbf{k}',\omega')\rangle$. Only the loop diagrams (left) contribute to this average. The vacuum diagrams (right) do not contribute.}
    \label{fig:FirstOrderDiags}
\end{figure}

\begin{figure}
    \centering
    \includegraphics[width=\textwidth]{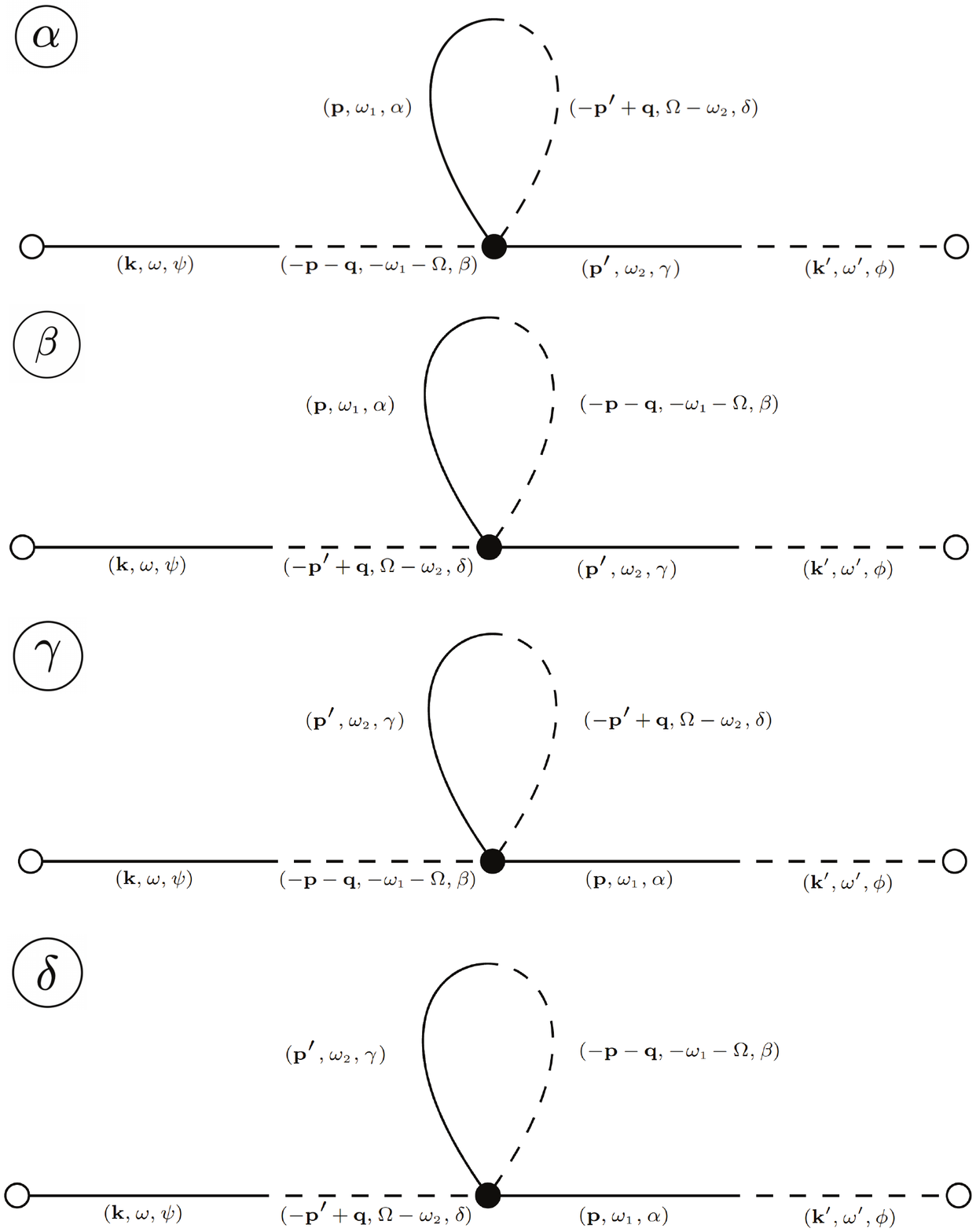}
    \caption{The non-zero first order diagrams whose sum represents the quantity (\ref{eq:FirstOrdAvg}) after applying Wick's theorem. The notation is the same as in Fig. \ref{fig:ZeroDiag}. The vertices (black dots) contribute a factor of $M_{\alpha\beta\gamma\delta}^{kl} (\mathbf{q},\Omega) p_k p_l'$.}
    \label{fig:FirstDiag}
\end{figure}

Compared to the zero-order diagram, the first order diagrams gain an interaction vertex. All possible types of first order diagram are shown in Fig. \ref{fig:FirstOrderDiags}. 
Note that vacuum diagrams (where the loops are separated from the external lines) do not contribute, as they are cancelled by the normalisation. This is referred to as the ``linked cluster theorem''\cite{AltlandSimons}. Extensions to higher order would involve two or more loops and interaction vertices.
Since any combinations involving two $\mathbf{A}$s vanish (see (\ref{eq:S0Matrix})), we need only consider diagrams where $\mathbf{A}$ and $\tilde{\mathbf{A}}$ connect; all unique non-zero combinations to first order are given in Fig. \ref{fig:FirstDiag}. 
Let us begin by copmuting the contribution to (\ref{eq:AAtilde}) from diagram \circled{$\alpha$}. Following our rules, we may write down
\begin{align}
\begin{split}
\label{eq:AlphaContribution}
       \text{\circled{$\alpha$}} =&\langle A_\psi (\mathbf{k},\omega) \tilde{A}_\beta (-\mathbf{p}-\mathbf{q},-\omega_1-\Omega)\rangle \langle A_\gamma (\mathbf{p}',\omega_2)\tilde{A}_\phi (\mathbf{k'},\omega') \rangle \\
       & \hspace{5cm} \ \langle A_\alpha (\mathbf{p},\omega_1) \tilde{\mathbf{A}}_\delta (\mathbf{q}-\mathbf{p}',\Omega-\omega_2) \rangle \\
    =& -\frac{i}{2} \int d \mathbf{p} d \mathbf{p}' d \mathbf{q}  d\omega_1 d\omega_2 d\Omega \ p_k p_l'  M_{\alpha\psi\phi\alpha}^{kl}(\mathbf{q},\Omega) \delta(\omega_1+\Omega-\omega_2)  \delta(\omega-\omega_1-\Omega) \\
    & \qquad \ \ \  \delta(\omega_2+\omega')\delta (\mathbf{q}-\mathbf{p}'+\mathbf{p}) \delta(\mathbf{k}-\mathbf{p}-\mathbf{q})\delta(\mathbf{p}'+\mathbf{k}') G_0(\mathbf{p},\omega_1) G_0(\mathbf{k},\omega) G_0(\mathbf{k}',\omega'),
\end{split}
\end{align}
where $G_0(\mathbf{k},\omega)$ is
\begin{align}
    G_0(\mathbf{k},\omega) =(i\omega+\eta k^2)^{-1}.
\end{align}
Taking the integrals over $\mathbf{p}$, $\mathbf{p}'$, $\omega_1$ and $\omega_2$  we find
\begin{align}
    I_\alpha &= -\frac{i}{2} \delta(\mathbf{k}+\mathbf{k}')\delta(\omega+\omega') G_0^2(\mathbf{k},\omega)\int d \mathbf{q} d\Omega \ G_0(\mathbf{k}-\mathbf{q},\omega-\Omega)(\mathbf{k}-\mathbf{q})_k k_l M_{\alpha\psi\phi\alpha}^{kl}(\mathbf{q},\Omega),
\end{align}
where
\begin{align}
\label{eq:Mappa}
    M_{\alpha\psi\phi\alpha}^{kl}(\mathbf{q}) =\big(
    \delta_{\phi\psi}\delta_{il}\delta_{jk}
    - 
    \delta_{i\phi}\delta_{l\psi}\delta_{jk}
    \big) \gamma(\mathbf{q},\Omega) \sigma_{ij}(\mathbf{q}).
\end{align}
Let us consider the contributions arising from 
the two terms in (\ref{eq:Mappa}) separately. The contribution from the first term is
\begin{align}
\label{eq:AlphaDiagCont1}
     \delta_{\phi\psi}\delta_{il}\delta_{jk} (\mathbf{k}-\mathbf{q})_k k_l \ \gamma(\mathbf{q},\Omega)\sigma_{ij}(\mathbf{q}) &= \gamma(\mathbf{q},\Omega) \big(\mathbf{k}\cdot \sigma(\mathbf{q})\cdot(\mathbf{k}-\mathbf{q})\big)\delta_{\phi\psi}\\
    &=\gamma(\mathbf{q},\Omega) \bigg(k^2-\frac{(\mathbf{q}\cdot\mathbf{k})^2}{q^2}\bigg)\delta_{\phi\psi},
\end{align}
while the second term is
\begin{align}
\label{eq:AlphaDiagCont2}
       & \delta_{i\phi}\delta_{l\psi}\delta_{jk} (\mathbf{k}-\mathbf{q})_k k_l \ \gamma(\mathbf{q},\Omega)\sigma_{ij}(\mathbf{q}) = \gamma(\mathbf{q},\Omega)k_\psi \sigma_{\phi j}(\mathbf{q}) (\mathbf{k}-\mathbf{q})_j . 
\end{align}
The difference between these two terms is that (\ref{eq:AlphaDiagCont1}) only includes diagonal terms, whereas (\ref{eq:AlphaDiagCont2}) has off-diagonal elements. %
These corrections are for the response of the magnetic vector potential, not the magnetic field, so they will be used in (\ref{eq:GAtoGB}) to find $\mathcal{G}_{\alpha\beta}(\mathbf{k},\mathbf{k}',\omega,\omega')$. Since (\ref{eq:GAtoGB}) includes terms of the form $\varepsilon_{\alpha l\phi}\varepsilon_{\beta n\psi} k_n' k_l$ (which represent cross products with $\mathbf{k}$), (\ref{eq:AlphaDiagCont2}) cannot contribute to the magnetic field's Green function, as its cross product with $\mathbf{k}$ is zero. This is true in general, so any of the diagrams for which we have "free" components of $k$ will vanish.

Let us now consider the contribution from the \circled{$\beta$} diagram. Here we have
\begin{align}
\begin{split}
\label{eq:BetaContribution}
    \text{\circled{$\beta$}} =-\frac{i}{2} \int d \mathbf{p} d \mathbf{p}' d \mathbf{q} d\omega_1 d\omega_2 &d\Omega \ p_k p_l' M_{\alpha\alpha\psi\phi}^{kl}(\mathbf{q},\Omega) \delta(-\Omega)\delta(\omega-\omega_2+\Omega)\delta(\omega_2+\omega')\\ &\delta (-\mathbf{q})\delta(\mathbf{k}-\mathbf{p}'+\mathbf{q})\delta(\mathbf{p}'+\mathbf{k}') G_0(\mathbf{p},\omega_1) G_0(\mathbf{k},\omega) G_0(\mathbf{k}',\omega'),
\end{split}
\end{align}
where 
\begin{align}
    M_{\alpha\alpha\psi\phi}^{kl}(\mathbf{q},\Omega)= -2\gamma(\mathbf{q},\Omega)(\delta_{l\phi}\sigma_{\psi k}- \delta_{\psi\phi} \sigma_{lk}),
\end{align}
so
\begin{align}
\begin{split}
\label{eq:BetaContribution2}
     \text{\circled{$\beta$}} =+i \int d \mathbf{p} d \mathbf{q}d\omega_1 d\Omega & \ p_k k_l' \gamma(\mathbf{q},\Omega)(\delta_{l\phi}\sigma_{\psi k}- \delta_{\psi\phi} \sigma_{lk}) \delta(\omega+\omega'+\Omega)\delta(-\Omega)\\
     &\delta (-\mathbf{q})\delta(\mathbf{k}+\mathbf{k}'+\mathbf{q}) G_0(\mathbf{p},\omega_1) G_0(\mathbf{k},\omega) G_0(\mathbf{k}',\omega').
\end{split}
\end{align}
The entire diagram has a $k_l'$ dependence, so it will not contribute to the Green function for the induction equation. The same is true for the contribution from the \circled{$\gamma$} diagram.  
The \circled{$\delta$} diagram contributes
\begin{align}
\begin{split}
    \label{eq:DeltaContribution}
    \text{\circled{$\delta$}} = &-\frac{i}{2} G^2_0(\mathbf{k},\omega) \delta(\mathbf{k}+\mathbf{k}')\delta(\omega+\omega')\\ 
    &\int d \mathbf{q} d\Omega \ \gamma(\mathbf{q},\Omega) G_0(\mathbf{k}+\mathbf{q},\omega+\Omega) \left[\delta_{\psi\phi} \bigg(k^2 - \frac{(\mathbf{k}\cdot\mathbf{q})^2}{q^2}\bigg)  - k_k (\mathbf{k}+\mathbf{q})_l \delta_{k\psi}\sigma_{\phi l}\right].
\end{split}
\end{align}
Again, the off-diagonal term proportional to $k_k$ does not contribute to the induction equation Green function. Overall, the interaction contribution to the vector potential's Green function is
\begin{align}
    G_{\psi\phi}^{int}(\mathbf{k},\mathbf{k}',\omega,\omega')=-i\langle\mathbf{A}_\psi(\mathbf{k},\omega)\tilde{\mathbf{A}}_\phi(\mathbf{k}',\omega')\rangle_{\text{int}} = -i\left( \text{\circled{$\alpha$}} +  \text{\circled{$\delta$}}\right).
\end{align}
Using (\ref{eq:GAtoGB}), (\ref{eq:AlphaContribution}) and (\ref{eq:DeltaContribution}), we finally find the induction equation's Green function to first order to be
\begin{align}
\begin{split}
    \mathcal{G}_{\alpha\beta} (\mathbf{k},\mathbf{k}',\omega,\omega')=&\frac{(2\pi)^4}{k^2} \delta(\mathbf{k}+\mathbf{k}')\delta(\omega+\omega')\varepsilon_{\alpha l\phi}\varepsilon_{\beta n\psi} k_n k_l \delta_{\phi\psi} \bigg[ \frac{-i^2}{i\omega + \eta k^2} \\
    &+\frac{i^2}{2(i\omega + \eta k^2)^2}\int \dbar \mathbf{q} \dbar\Omega \frac{\gamma(\mathbf{q},\Omega)}{i(\omega-\Omega)+\eta(\mathbf{k}-\mathbf{q})^2} \bigg(k^2-\frac{(\mathbf{q}\cdot\mathbf{k})^2}{q^2}\bigg)\\
     &+\frac{i^2}{2(i\omega + \eta k^2)^2}\int \dbar \mathbf{q} \dbar\Omega \frac{\gamma(\mathbf{q},\Omega)}{i(\omega+\Omega)+\eta(\mathbf{k}+\mathbf{q})^2} \bigg(k^2-\frac{(\mathbf{q}\cdot\mathbf{k})^2}{q^2}\bigg)\bigg].
\end{split}
\end{align}
Making use of $\varepsilon_{\beta n\psi}\varepsilon_{\alpha l\phi} k_n k_l \delta_{\phi\psi} = k^2 \sigma_{\alpha\beta}(k)$, we find that this is equivalent to
\begin{align}
\label{eq:GreensIndEq}
    \begin{split}
    \mathcal{G}_{\alpha\beta} (\mathbf{k},\mathbf{k}',\omega,\omega')  =(2\pi)^4\delta(\mathbf{k}+\mathbf{k}') \delta(\omega+\omega') \frac{\sigma_{\alpha\beta}(k)}{i\omega + \eta k^2} \bigg[ 1-\frac{1}{2}\int \dbar \mathbf{q} \dbar \Omega \frac{k^2-q^{-2}(\mathbf{q}\cdot\mathbf{k})^2}{i\omega + \eta k^2} &\\
    \left( \frac{\gamma(\mathbf{q},\Omega)}{i(\omega-\Omega)+\eta(\mathbf{k}-\mathbf{q})^2}+\frac{\gamma(\mathbf{q},\Omega)}{i(\omega+\Omega)+\eta(\mathbf{k}+\mathbf{q})^2}\right) \bigg]&\\
    = (2\pi)^4\delta(\mathbf{k}+\mathbf{k}') \delta(\omega+\omega') \sigma_{\alpha\beta}(k) g(k,\omega),\hspace{3.7cm}&
    \end{split}
\end{align}
where $g(k,\omega)$ is defined as
\begin{align}
\begin{split}
\label{eq:g(w,k)}
    g(k,\omega) = \frac{1}{i\omega + \eta k^2} \bigg[ 1-\frac{1}{2}\int \dbar \mathbf{q} \dbar\Omega \frac{k^2-q^{-2}(\mathbf{q}\cdot\mathbf{k})^2}{i\omega + \eta k^2}\left(  \frac{\gamma(\mathbf{q},\Omega)}{i(\omega-\Omega)+\eta(\mathbf{k}-\mathbf{q})^2}+ \right. \qquad &\\
    \left. \frac{\gamma(\mathbf{q},\Omega)}{i(\omega+\Omega)+\eta(\mathbf{k}+\mathbf{q})^2}\right)\bigg]& .
\end{split}
\end{align}
Having found the magnetic field's Green function to first order for a general correlation function, we may now ask how the fluid flow's turbulent behaviour affects physical properties of the system, such as the magnetic diffusivity.

\section{Turbulent diffusivity}\label{sec:DiffCalc}

Let us begin by considering how the diffusivity is defined. Generally, it may be thought of as the coefficient of the induction equation's Laplacian term in physical space, or the $k^2$ term in Fourier space. In the zero-order case, the Green function is
\begin{align}
    \mathcal{G}^0_{\alpha\beta}(\mathbf{k},\mathbf{k}',\omega,\omega') = (2\pi)^4\delta(\mathbf{k}+\mathbf{k}')\delta(\omega+\omega') \sigma_{\alpha\beta}(k) g_0(k,\omega).
\end{align}
The delta functions reflect the spatial and temporal translation invariance, while $\sigma_{\alpha\beta}(k)$ guarantees the magnetic field's solenoidality. Any system-specific information is contained in $g_0(k,\omega)$. Let us consider its inverse
\begin{align}
    g_0^{-1}(k,\omega) = i\omega + \eta k^2,
\end{align}
where $\eta$ is the molecular (zero-order) diffusion coefficient. Defining $\omega_0^*(k)$ as the solution to $g_0^{-1}(\omega_0^*,k)=0$, we may write $\omega_0^* = i\eta k^2 $, from which we can see that
\begin{align}
\label{eq:Eta0Def}
    \eta = \frac{1}{2i} \frac{\partial^2 \omega_0^*}{\partial k^2}\bigg|_{k=0}
\end{align}
is a valid definition for the diffusion coefficient. 
Note that the solutions $\omega_0^*$ are the poles of $g_0(k,\omega)$, which are the reciprocals of the decay times. 

To compute the new diffusivity for the induction equation, we write the Green function as
\begin{align}
\label{eq:Diffg(w,k)Def}
    g(k,\omega) = i\omega + \eta_{\text{eff}}k^2 + \dots,
\end{align}
where $\eta_{\text{eff}}$ is a modified diffusivity and `$\dots$' represents more complex terms that might contribute to the response. 
To find an expression for $\eta_{\text{eff}}$, we require the inverse of (\ref{eq:GreensIndEq}).
Note that $g(\omega,k)$, defined in (\ref{eq:g(w,k)}), may be written in terms of $g_0(k,\omega)$ as
\begin{align}
    g(k,\omega) = g_0(k,\omega) \left[1 - g_0(k,\omega)f(k,\omega)\right],
\end{align}
where
\begin{align}
    f(k,\omega) = \frac{1}{2}\int \dbar \mathbf{q} \dbar\Omega \left(k^2-\frac{(\mathbf{q}\cdot\mathbf{k})^2}{q^2}\right)\left[ \frac{\gamma(\mathbf{q},\Omega)}{i(\omega-\Omega)+\eta(\mathbf{k}-\mathbf{q})^2}+ \frac{\gamma(\mathbf{q},\Omega)}{i(\omega+\Omega)+\eta(\mathbf{k}+\mathbf{q})^2}\right].
\end{align}
Assuming that the first order correction is small, we may write
\begin{align}
\begin{split}
\label{eq:InvGreensIndEq}
    g^{-1} (k,\omega) &= 
     g_0^{-1}(k,\omega) + f(k,\omega),
\end{split}
\end{align}
Let us now look for the poles of $g(k,\omega)$ by solving $g^{-1}(\omega^*, k)=0$ for $\omega^*(k)$. From (\ref{eq:InvGreensIndEq}) we see that they must occur when
\begin{align}
\begin{split}
\label{eq:PolesGB}
    \omega^*(k) = i \eta k^2 +if(k,\omega^*).
\end{split}
\end{align}
Note that $f(k,\omega^*)$ is of the form $k^2F(k,\omega^*)$,
where
\begin{align}
\begin{split}
F(k,\omega^*) =\frac{1}{2} \int \dbar \mathbf{q} \dbar\Omega \left(1-\cos^2\theta\right)& \left[  \frac{\gamma(\mathbf{q},\Omega)}{i(\omega^*-\Omega)+\eta(\mathbf{k}-\mathbf{q})^2}+ \frac{\gamma(\mathbf{q},\Omega)}{i(\omega^*+\Omega)+\eta(\mathbf{k}+\mathbf{q})^2}\right]
\end{split}
\end{align}
and the angle between $\mathbf{k}$ and $\mathbf{q}$ has been labelled $\theta$. In analogy to  (\ref{eq:Eta0Def}), we may find the new effective diffusivity by considering
\begin{align}
   \eta_{\text{eff}} = \left.\frac{1}{2i} \frac{\partial^2\omega^*}{\partial k^2}\right\lvert_{k=0} = \eta + F(0,\omega^*(0)).
\end{align}
The first term is of course the zero-order molecular diffusivity, while the second term constitutes the contribution from the turbulent diffusivity, $\eta_t$.
Given that $\omega^*(0) = 0$, we find
\begin{align}
    \eta_t = \int \dbar\mathbf{q} \dbar \Omega \ \gamma(\mathbf{q},\Omega)\left(1-\cos^2\theta\right) \frac{\eta q^2}{\Omega^2+\eta^2 q^4}. 
\end{align}
Assuming that $\gamma(\mathbf{q},\Omega)$ only depends on the magnitude $\lvert \mathbf{q} \lvert = q$ (isotropic approximation), we may take the integral over the solid angle, and find
\begin{align}
\label{eq:EtaT}
    \eta_t = \frac{4}{3} \eta \int dq d\Omega \frac{q^4}{(2\pi)^3}  \frac{\gamma(q,\Omega)}{\Omega^2+\eta^2 q^4}. 
\end{align}
We have finally found an expression for the magnetic diffusivity in terms of the velocity's correlation function. Expression (\ref{eq:EtaT}) gives us a quantitative estimate of the effects of turbulent diffusion on the 
magnetic turbulent diffusivity.

It is worthwhile pointing out that there are instances where the expression for the diffusivity cannot be valid on all lengthscales, though these are not physically realistic scenarios. Consider for example the simple case where $\gamma(q,\Omega)$ is a constant. In this case, the integral (\ref{eq:EtaT}) diverges. This does not mean that there is an issue with expression (\ref{eq:EtaT}), rather that this expression cannot be applicable to all scales, and there must be a wavenumber at which the integral is cut off. At very large wavenumbers the theory cannot possibly capture the dominant physics, as we would be considering lengthscales smaller than the size of an atom (and our theory is not quantum mechanical).

It is insightful to compare our result to those of Kazantsev\cite{Kazantsev} and Moffatt\cite{Moffatt78}, who studied the same problem by different means. We delve into this comparison in the next section.

\section{Comparison to literature and physical implications}

Our expression for the turbulent diffusivity under the assumption of isotropic turbulence (\ref{eq:EtaT})  may be compared to well-established literature.
In particular, we will compare our result to those of Moffatt\cite{Moffatt78} and Kazantsev \cite{Kazantsev}, who provided estimates of the turbulent diffusivity that might be expected in the case of isotropic turbulence in a conductive fluid.
Moffatt computed an expression for the turbulent diffusivity via the mean-field theory framework developed by Steenbeck, Krause and R\"adler\cite{MeanFieldTheory}, and found
\begin{align}
    \label{eq:EtaTMoffatt}
    \eta_t = \frac{2}{3} \eta \int dq d\Omega \ \frac{q^2 E(q,\Omega)}{\Omega^2 +\eta^2 q^4}, 
\end{align}
where $E(q,\Omega)$ is defined as the energy spectrum function, such that
\begin{align}
    \int dq d\Omega \ E(q,\Omega)= \frac{1}{2}\langle\mathbf{u}^2\rangle.
\end{align} 
Physically, $\rho \langle \mathbf{u}^2\rangle/2$ is the kinetic energy density, where $\rho$ is the fluid's density.
Taking into account Moffatt's different definition of the Fourier transform, we may relate $E(q,\Omega)$ to $\gamma(q,\Omega)$ via
\begin{align}
\label{eq:EtoGamma}
    E(q,\Omega) = \frac{2 q^2 \gamma(q,\Omega)}{(2\pi)^3}.
\end{align}
Plugging (\ref{eq:EtoGamma}) into (\ref{eq:EtaTMoffatt}) reveals that our estimate of $\eta_t$ is identical to that proposed by Moffatt using mean-field theory. The agreement between our results and Moffatt's at this stage is perhaps not too surprising. The first-order smoothing approximation used in mean-field theory is analogous to the Born approximation in scattering theory\cite{Moffatt78}. Our first order expression for the Green function (\ref{eq:GreensIndEq}) is exactly the Born approximation in the language of scattering theory.
This agreement also lends credence to the assumption of a Gaussian probability distribution for the velocity (at least in studying this specific problem), as Moffatt does not make any assumptions about the velocity's statistics. 
Though it may appear more laboured, the advantage of our method is that it can be extended to include an infinite number of terms in the perturbation series by exploiting the field theory formalism.

Let us now compare our results to Kazantsev's study. Kazantsev developed a model for the turbulent motion of a conductive fluid, assuming that the fluid flow has a Gaussian distribution function and is delta-correlated in time. This problem's set-up is very similar to our own, though Kazantsev works directly with the magnetic field (rather than the magnetic vector potential), and imposes solenoidality through the enforcement of initial conditions. By establishing a recursive relation for the magnetic field's Green function, he was able to extract an expression for the turbulent diffusivity contribution of the form
\begin{align}
\label{eq:EtaTKazantsev}
    \eta_t = \frac{1}{3} \int dq\ q^2 \frac{\gamma(q)}{(2\pi)^3},
\end{align}
where $\gamma(q)$ is now purely a function of wavevector. This result may be readily derived from (\ref{eq:EtaT}) by taking the velocity to be delta-correlated in time.
Moffatt and Kazantsev's estimates may be thought of as being valid in different limits of the magnetic Reynolds number, $Rm$.  Kazantsev's assumption of instantaneous time correlation requires the correlation time $\tau$ to be small compared to the timescale $\ell/\eta$, where $\ell$ is the velocity's correlation length. The inequality $\tau \ll \ell/\eta$ may be recast as $Rm\gg1$, suggesting that Kazantsev's result is the high $Rm$ limit of Moffatt's more general result (\ref{eq:EtaTMoffatt}). 
The magnetic Reynolds number here refers specifically to the fluctuating part of the velocity.

The geodynamo is thought to operate at a global $Rm\sim 1000$ \cite{Christensen04}, though the scale separation makes it more appropriate to consider a local $Rm$ when considering turbulent processes. If the turbulent flow's lengthscale were to correspond to the dissipation lengthscale, $Rm$ could be as low as $Rm\sim1$ \cite{DavidsonTurbulence}. Holdenried-Chernoff and Buffett \cite{BuffettHoldenried22} found that a local $Rm$ based on the eddy dissipation lengthscale might be an order of magnitude smaller than the large scale $Rm$, so we might expect $Rm<100$ at the lengthscales relevant to the turbulent flow. These are quite moderate values, so if we wanted to consider the turbulent diffusivity in the context of the geodynamo we should probably use the more general expression (\ref{eq:EtaT}). 
However, we do not have any information about the form of $\gamma(q,\Omega)$ that would be expected for velocity correlations in the Earth. 
As an illustrative example, let us consider the simple case of exponentially correlated velocities in both space and time, such that
\begin{align}
    \langle \mathbf{u}(\mathbf{x},t) \cdot \mathbf{u}(\mathbf{x}',t') \rangle = u_{\text{rms}}^2 e^{-\lvert \mathbf{x} - \mathbf{x}'\lvert/\ell} e^{-\lvert t - t'\lvert/\tau},
\end{align}
where $u_{\text{rms}}$ is the average root mean square velocity in the outer core.
In conjunction with (\ref{eq:EtaT}), this correlation function yields a turbulent diffusivity
\begin{align}
    \eta_t = \frac{1}{3} u_{\text{rms}}^2 \tau \left(1+\sqrt{\frac{\eta \tau}{\ell^2}} \right)^{-2}.
\end{align}
The last term may be thought of in terms of a local turbulent $Rm^t$, defined by the correlation time and lengthscales
\begin{align}
    Rm^t = \frac{\ell^2}{\eta \tau}.
\end{align}
As we either take $\tau\to0$ (short correlation times) or $\eta \to 0$ (the high $Rm^t$ limit), this expression happens to correspond to the relation derived by Holdenried-Chernoff and Buffett \cite{BuffettHoldenried22} using mean-field theory, namely
\begin{align}
    \label{eq:EtaTUrms}
    \eta_t = \frac{1}{3} u_{\text{rms}}^2 \tau.
\end{align}
 Using (\ref{eq:EtaTUrms}) and estimates of the magnetic dipole's correlation times from paleomagnetic records, Holdenried-Chernoff and Buffett were able to compute an estimate for the average flow velocities expected in the outer core, and found that they were consistent with measurements of core surface velocities. 
 Accounting for the effects of finite $Rm^t\gtrsim 1$, we might expect the estimate of $u_{\text{rms}}$ to increase at most by a factor of 2.

\section{Extension to all orders} \label{sec:AllOrders}

In sections \ref{sec:FirstOrderResponse}\ref{sec:DiagsIntCorr} and \ref{sec:DiffCalc} we computed the first order correction to the propagator and to the diffusion coefficient.
While this is a good first step towards understanding the effects of turbulence on the magnetic field's decay rate, it is only an approximation to the true turbulent diffusivity. Crucially, the results we have obtained so far require the amplitude of the velocity correlations, $u_{\text{rms}}$, to be small compared to the molecular diffusivity.
To compute the full propagator we need to not only include the one vertex diagrams from the previous section, but also those with two vertices, three vertices and so on.
We may improve our first-order estimate of the propagator by using a technique pioneered by Dyson\cite{Dyson49} to formally including all diagrams. 

\begin{figure}
    \centering
    \includegraphics[width=\textwidth]{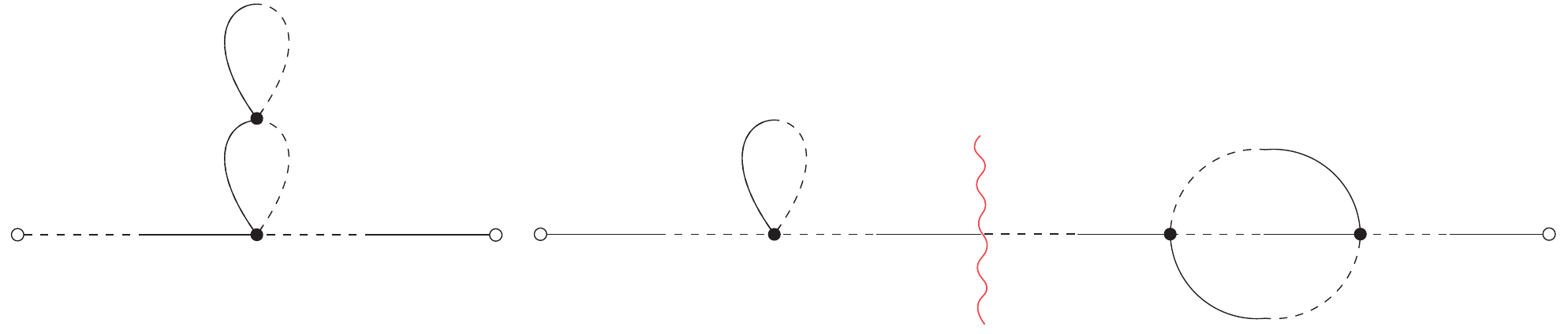}
    \caption{The diagram on the left is 1PI, as cutting any internal line results in a physically meaningless diagram. The diagram on the right is $\overline{1\text{PI}}$, as it can be cut along the wavy red line to produce two 1PI diagrams.}
    \label{fig:1PIdiags}
\end{figure}

\begin{figure}
    \centering
    \includegraphics[width=\textwidth]{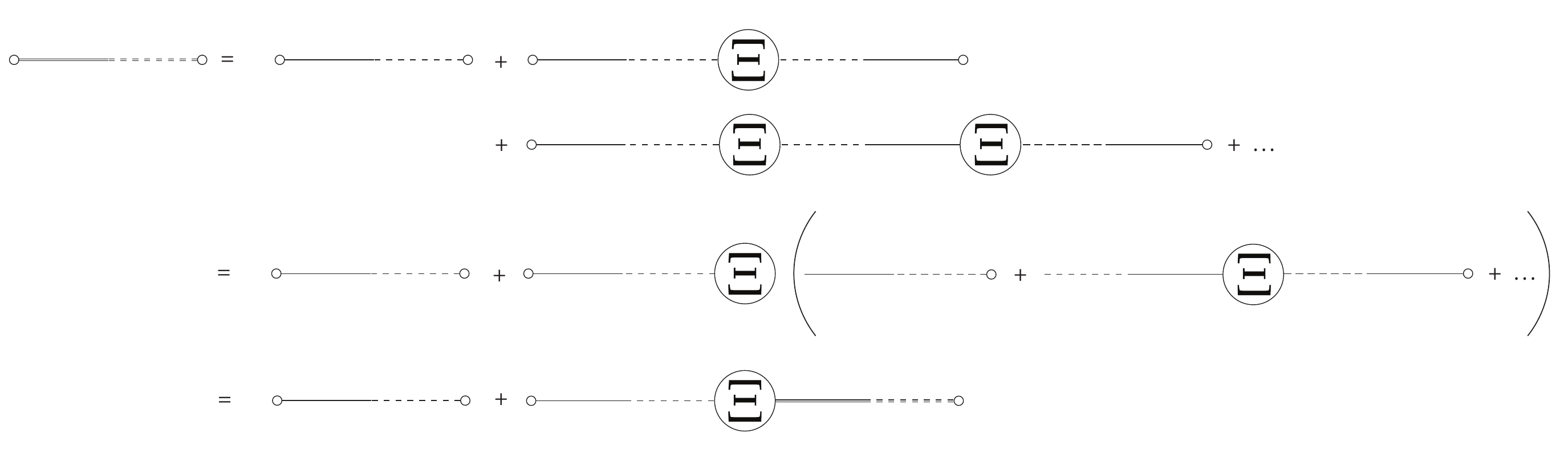}
    \caption{Diagrammatic illustration of the expansion for the quantity $\langle \mathbf{A}(\mathbf{k},\omega) \tilde{\mathbf{A}}(\mathbf{k}',\omega')\rangle$. The exact average is represented by double lines, where the solid and dashed lines indicate $\mathbf{A}$ and $\tilde{\mathbf{A}}$ respectively. The first diagram on the right-hand side is the purely diffusive, zero-order contribution to the average. Higher order contributions may be written in 1PI form, with each higher order having an additional $\Xi$ contribution ($\Xi$ is the sum of all 1PI diagrams). The exact propagator may be expressed as a geometric series, where each subsequent term includes an additional $\Xi$ contribution.}
    \label{fig:1PIAllOrders}
\end{figure}

Let us begin by separating the diagrams into two distinct types; those that are \textit{one particle irreducible} (1PI) and those that are not ($\overline{1\text{PI}}$). The latter type can be split up into two physically permissible diagrams by cutting one internal line, while in a 1PI diagram no internal line can be cut to produce a meaningful diagram. Figure \ref{fig:1PIdiags} illustrates the difference between 1PI and $\overline{1\text{PI}}$ diagrams.
From this example, it becomes clear that any $\overline{1\text{PI}}$ diagram may be built solely out of 1PI diagrams.
Denoting the sum of all 1PI diagrams $\boldsymbol{\Xi}$, we may reorder the expansion for the propagator (written up to first order in (\ref{eq:AAtilde})), so that it may be expressed diagrammatically as a geometric series, shown in Fig. \ref{fig:1PIAllOrders}.
This series includes every single diagram, and so is an exact expression for the propagator. Since it is written as a geometric series, we may sum the series to all orders.
Let us call $\Gamma_{\psi\phi}(\mathbf{k},\omega)$ the exact average, such that
\begin{align}
     \langle \mathbf{A}_\psi(\mathbf{k},\omega) \tilde{\mathbf{A}}_\phi(\mathbf{k}',\omega') \rangle = (2\pi)^4  \delta(\mathbf{k}+\mathbf{k}')\delta(\omega+\omega') i\Gamma_{\psi\phi} (\mathbf{k},\omega).
\end{align}
Note that the interaction cannot affect the spatial and time translation invariance of the system, which is why we have delta functions on the wavevectors and frequencies.
Fig. \ref{fig:1PIAllOrders} tells us that the magnetic potential's Green function ($i\Gamma_{\psi \phi}$) may be written to all orders as
\begin{align}
\begin{split}
\label{eq:AllOrdersGamma}
 \boldsymbol{\Gamma}(\mathbf{k},\omega) =&  \mathbf{G}(\mathbf{k},\omega) - \mathbf{G}(\mathbf{k},\omega) \cdot \boldsymbol{\Xi}(\mathbf{k},\omega) \cdot \mathbf{G}(\mathbf{k},\omega) + \mathbf{G}\cdot \boldsymbol{\Xi} \cdot \mathbf{G}\cdot \boldsymbol{\Xi} \cdot \mathbf{G} + \dots\\
 =&\mathbf{G}(\mathbf{k},\omega)\cdot \sum_{n=0}^\infty (-\boldsymbol{\Xi}(\mathbf{k},\omega) \cdot \mathbf{G}(\mathbf{k},\omega))^n = \mathbf{G}(\mathbf{k},\omega)\left(\mathbf{1}+\boldsymbol{\Xi}(\mathbf{k},\omega)\cdot\mathbf{G}(\mathbf{k},\omega)\right)^{-1},
\end{split}
\end{align}
where $\mathbf{1}$ is the identity matrix and the arguments have been omitted from the final term on the first line of (\ref{eq:AllOrdersGamma}) for notational clarity. The
matrix $\mathbf{G}(\mathbf{k},\omega)$ is a shorthand for the zero order contribution
\begin{align}
\mathbf{G} (\mathbf{k},\omega) = G_0(\mathbf{k},\omega) \mathbf{1}.
\end{align}
The solution for $\boldsymbol{\Gamma}(\mathbf{k},\omega)$ in (\ref{eq:AllOrdersGamma}) is formally exact. However, computing $\boldsymbol{\Xi}(\mathbf{k},\omega)$ exactly is an intractable problem, so we must resort to computing it perturbatively by writing it as a series where each term corresponds to a different 1PI diagram with an increasing number of vertices. In this way, while we are still neglecting an infinite number of terms that contribute to the propagator, we are also considering an infinite number of interactions that may be written in terms of specific 1PI representations.
As a simple example, let us consider the first order expansion of $\boldsymbol{\Xi}(\mathbf{k},\omega)$,
\begin{equation}
    \includegraphics[width=0.3\textwidth]{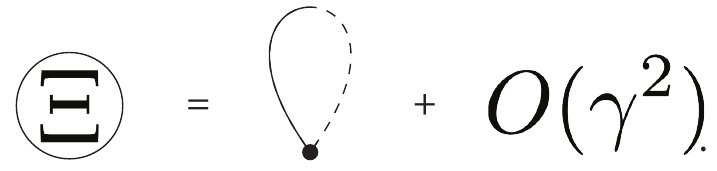}
\end{equation}
Through (\ref{eq:AllOrdersGamma}), we may include the loop's contribution to the propagator to all orders. Leaving the details of this calculation to Appendix \ref{app:AllLoopOrdDiffCalc}, we can compute an expression for the magnetic field's Green function to all orders in the loop diagram, and find
\begin{align}
\begin{split}
\label{eq:BsGreensFunctAllOrd}
\mathcal{G}_{\alpha \beta}(\mathbf{k},\omega) =& \frac{\sigma_{\alpha\beta}(\mathbf{k})}{G_0^{-1}(\mathbf{k},\omega)+\Xi(\mathbf{k},\omega)},
\end{split}
\end{align}
where 
\begin{align}
    \Xi (\mathbf{k},\omega) = \frac{1}{2} \int \dbar \mathbf{q} \dbar \Omega \ \gamma(\mathbf{q},\Omega) \big[G_0 (\mathbf{k}-\mathbf{q}, \omega-\Omega)+G_0 (\mathbf{k}+\mathbf{q},\omega+\Omega)\big] \bigg(k^2 - \frac{1}{q^2}(\mathbf{k}\cdot\mathbf{q})^2 \bigg).
\end{align}
Eq. (\ref{eq:BsGreensFunctAllOrd}) allows us to compute the influence of the additional loop diagrams on the effective diffusivity. 
Following the same procedure as in section \ref{sec:DiffCalc}, we find that, rather remarkably, the diffusivity remains unchanged compared to (\ref{eq:EtaT}).
In fact, this appears to be the calculation that Kazantsev\cite{Kazantsev} illustrates in Fig. 2 of his paper, with exactly the same outcome. 
While we have neglected an infinite number of contributions, the infinity of loop diagrams that we have considered agrees exactly with mean-field theory, which has been shown to be very successful \cite{Kraichnan76,Moffatt78} as long as the velocity's correlation time is small. This is suggestive, and it might hint that the neglected infinity of diagrams may not constitute an important contribution for short enough correlation times. Though we do not provide an answer here, the framework we have developed allows us to explore this question. 

Even though we have only expanded $\boldsymbol{\Xi}$ to first order, this scheme always contains more information than naive perturbation theory. This is because we are no longer restricting the velocity to have small amplitudes, as the all orders treatment illustrated in fig. (\ref{fig:1PIAllOrders}) takes into account all powers of $\boldsymbol{\Xi}$. The nature of the approximation made when $\boldsymbol{\Xi}$ is truncated at a given number of vertices is subtle, because different diagrams are relevant for different length and time scales.
Computing the contributions from the higher order diagrams is certainly worthwhile, to ascertain whether including them significantly affects the turbulent diffusivity. It will be interesting to see whether including these diagrams changes the propagator's properties (wavenumber and index structure) compared to the results given here.
The next order in $\boldsymbol{\Xi}$ must include all 1PI contributions with two vertices, namely
\begin{equation}
    \includegraphics[width=0.48\textwidth]{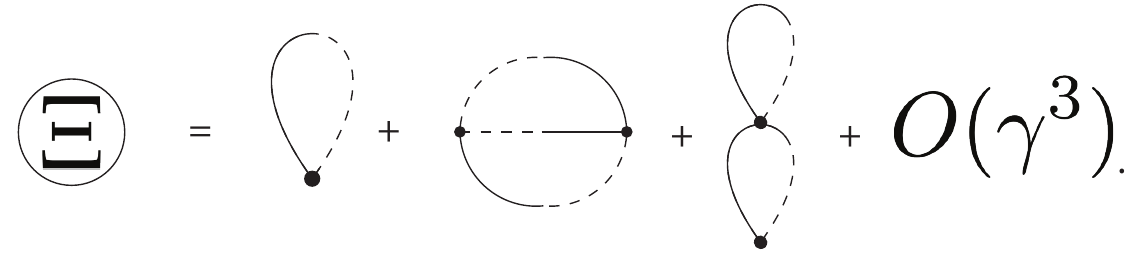}
\end{equation}
We leave this to future work, as the principle aim of this paper is to introduce the MSRJD framework to the problem of studying stochastic dynamos. 


\section{Conclusions} \label{sec:Conclusion}

Studying the induction equation in the framework of stochastic models seems to be a reasonable way of attempting to understand the long timescale behaviour of the Earth's magnetic field. A theoretical understanding of this behaviour can be coupled with statistical information extracted from paleomagnetic records to infer properties of the fluid flow underlying the geodynamo.

In this paper, we introduce methods commonly used in field-theory to create a framework that allows us to compute the ensemble averages of arbitrary observables of the magnetic field. By computing the magnetic field's average Green function, we find an expression for the turbulent magnetic diffusivity that is exactly consistent with the results of Moffatt's\cite{Moffatt78} mean-field theory estimate for all magnetic Reynolds numbers. The turbulent contribution to the diffusivity is positive (and thus enhances the magnetic field's decay), and depends on the spatio-temporal correlation function of the velocity.
Our results are also consistent with Kazantsev's\cite{Kazantsev} study when we restrict the velocity to be delta-correlated in time. The methods applied in this paper allow us to compute arbitrary observables of the magnetic field, and so can be used to extend Kazantsev's work.

Taking the high $Rm$ limit results in a relation between the turbulent diffusivity and the average root mean squared velocity in the bulk of the fluid. Making use of this expression along with an estimate of the turbulent diffusivity in the Earth from paleomagnetic constraints, Holdenried-Chernoff and Buffett \cite{BuffettHoldenried22} found an upper bound on the possible average fluid velocities in the outer core. This estimate is consistent with measured core surface flows, suggesting that although both mean-field theory and our approach make a number of simplifying assumptions compared to the true Earth system, they should at least be capturing the correct order of magnitude of the turbulent effects.

In principle, we could restrict our calculation to spherical geometry by imposing the velocity's correlations to go to zero outside the sphere's radius. While the change in geometry would most probably have some quantifiable impact on the expressions we have derived, it should not drastically alter the system's physics (and consequently the expressions' qualitative structure), though it would restrict the set of velocities that we should consider when averaging. 

In this study we have looked specifically at the diffusivity.
We could equally have considered the stochastic velocity's effect on other processes, such as inductive processes akin to the alpha effect. These come in as terms that are independent of $k$ in the pole. In order to examine these inductive effects we would need to see how the coefficients of $\mathbf{B}$ terms are altered (in the purely diffusive case these coefficients are of course zero). The particular set-up that we have chosen does not support an alpha effect, as the velocity's correlations are symmetric upon reflection. The alpha effect is intrinsically linked to the fluid's helicity, which requires the system to have a lower symmetry than we've imposed. Our model may be easily extended to account for this by using the more general velocity correlation function suggested by Moffatt (eq. (7.56) in \cite{Moffatt78}).

Overall, we believe that applying field-theory methods to the stochastic induction equation has great potential for addressing a number of interesting questions. For example, if we were to introduce a mean fluid velocity that acts as a kinematic dynamo, we could ask how the stochastic motions affect the fluid's ability to induce magnetic field, and investigate whether turbulence is beneficial or detrimental to sustaining the dynamo.
Ultimately, the greatest flaw in our model is the need to assume the form of the velocity's probability distribution. The final goal in stochastic geodynamo studies should be to develop a fully self-consistent model that accounts for the Navier-Stokes equation as well as the induction equation. 
\\


\noindent \textbf{Data Availability:} This article has no additional data.

\noindent \textbf{Author Contributions:} D.H.C.: conceptualization, formal analysis, funding acquisition, investigation, methodology, validation, visualization, writing – original draft,  writing – review \& editing. D.A.K.: conceptualization, formal analysis, methodology, validation, writing – review \& editing. B.A.B: conceptualization, funding acquisition, resources, supervision, validation, writing – review \& editing.

\noindent \textbf{Declaration:} We declare that we have no competing interests.

\noindent \textbf{Acknowledgments:} This work was supported by a Swiss National Science Foundation Early Postdoc.Mobility fellowship to DHC (P2EZP2\_199996) and by a grant (EAR-2214244) to BAB from the US National Science Foundation. D.A.K. was supported by the Simons Investigator Grant \#291825 from the Simons Foundation. We wish to thank Chris Scullard for helpful comments.


\appendix
\section{Fourier transform of interaction term}\label{app:ToFourier}

The general, incompressible form of the probability distribution for the velocity in Fourier space is
\begin{align}
\label{eq:ProbDistU}
    p[\mathbf{u}] = \exp\bigg[-\frac{1}{2} \int \dbar \mathbf{q}\ \dbar \mathbf{q}'\ d\omega\ d\omega'\ u_j (\mathbf{q},\omega) C^{-1}_{jk} (\mathbf{q},\mathbf{q}', \omega,\omega') u_k (\mathbf{q}',\omega') \bigg],
\end{align}
 where $C^{-1}_{jk}(\mathbf{q},\mathbf{q}', t,t')$ is the inverse of the correlation
 \begin{align}
 \label{eq:UCorrFourierApp}
     \langle u_j(\mathbf{q},\omega) u_k(\mathbf{q}', \omega')\rangle = C_{jk}(\mathbf{q},\mathbf{q}', \omega,\omega') = (2\pi)^4 \gamma(\mathbf{q},\omega) \delta(\mathbf{q}+\mathbf{q}')\delta(\omega+\omega') \sigma_{jk}(\mathbf{q}).
 \end{align}
 The function $\gamma(\mathbf{q},\omega)$ represents any spatial and temporal structure of the noise and $\sigma_{jk}(\mathbf{q})$ is defined in (\ref{eq:SigmaDef}).
To make use of  (\ref{eq:UCorrFourierApp}), we want to express the probability distribution $p[\mathbf{A},\tilde{\mathbf{A}}]$ in terms of Fourier variables. A simple way of doing this is to write each of the variables appearing in the definitions (\ref{eq:Sdefs}) in terms of their inverse Fourier transform. Let us consider how this works for the interaction term $S_{\text{int}}$, restricting our attention to the spatial Fourier transform. Time dependence is kept implicit for clarity, though the Fourier transform in time follows in the same manner. From (\ref{eq:Sdefs}) we have
\begin{align}
\begin{split}
S_{\text{int}}  
= \int d \mathbf{x}\ d \mathbf{x}' \int \dbar \mathbf{q}_{1\to 6} \ e^{i\mathbf{q}_{1\to 3} \cdot \mathbf{x}} e^{i\mathbf{q}_{4 \to 6} \cdot \mathbf{x}'} \varepsilon_{jab} \varepsilon_{kcd}(-i\mathbf{q}_1 \times \mathbf{A}(\mathbf{q}_1))_a \tilde{A}_b(\mathbf{q}_2) C_{jk}(\mathbf{q}_3,\mathbf{q}_4) &\\
(-i\mathbf{q}_5\times \mathbf{A}(\mathbf{q}_5))_c \tilde{A}_d(\mathbf{q}_6)&\\
=\ \ -(2\pi)^3\int d \mathbf{x}\ d \mathbf{x}' \int \dbar \mathbf{q}_{1\to 6} e^{i\mathbf{q}_{1\to 3} \cdot \mathbf{x}} e^{i\mathbf{q}_{4 \to 6} \cdot \mathbf{x}'} \varepsilon_{jab} \varepsilon_{kcd}\gamma(\mathbf{q}_3) \sigma_{jk}(\mathbf{q}_3)\delta(\mathbf{q}_3+\mathbf{q}_4) \hspace{0.5cm} &\\
(\mathbf{q}_1 \times \mathbf{A}(\mathbf{q}_1))_a \tilde{A}_b(\mathbf{q}_2) (\mathbf{q}_5\times \mathbf{A}(\mathbf{q}_5))_c \tilde{A}_d(\mathbf{q}_6)&\\
= \ \ \int d \mathbf{x}\ d \mathbf{x}' \int \dbar \mathbf{q}_{1\to 3}\dbar \mathbf{q}_{5\to 6} e^{i\mathbf{q}_{1\to 3} \cdot \mathbf{x}} e^{i(\mathbf{q}_{5}+\mathbf{q}_6-\mathbf{q}_3) \cdot \mathbf{x}'} \varepsilon_{jab} \varepsilon_{kcd}\gamma(\mathbf{q}_3) \sigma_{jk}(\mathbf{q}_3) \hspace{1.2cm} &\\ 
(\mathbf{q}_1 \times \mathbf{A}(\mathbf{q}_1))_a  \tilde{A}_b(\mathbf{q}_2) (\mathbf{q}_5\times \mathbf{A}(\mathbf{q}_5))_c \tilde{A}_d(\mathbf{q}_6)&.
\end{split}
\end{align}
Note that we may write
\begin{align}
\begin{split}
    &\int d\mathbf{x} e^{i(\mathbf{q}_{1}+\mathbf{q}_{2}+\mathbf{q}_{3}) \cdot \mathbf{x}} = (2\pi)^3\delta(\mathbf{q}_1+\mathbf{q}_2+\mathbf{q}_3), \text{ and }\\
    &\int d\mathbf{x}' e^{i(\mathbf{q}_5+\mathbf{q}_6-\mathbf{q}_3)\cdot\mathbf{x}'} =(2\pi)^3\delta(\mathbf{q}_5+\mathbf{q}_6-\mathbf{q}_3).
\end{split}
\end{align}
Using this property, and taking the integrals over $\mathbf{q}_3$ and $\mathbf{q}_6$, we obtain
\begin{align}
\label{eq:InteractionDerivation1}
\begin{split}
S_{\text{int}} = - \int \dbar \mathbf{q}_{1} \dbar \mathbf{q}_2 \dbar \mathbf{q}_5 \  \varepsilon_{jab} \varepsilon_{kcd}\gamma(-\mathbf{q}_1-\mathbf{q}_2) \sigma_{jk}(-\mathbf{q}_1-\mathbf{q}_2) (\mathbf{q}_1 \times \mathbf{A}(\mathbf{q}_1))_a \tilde{A}_b(\mathbf{q}_2)  \hspace{1cm}&\\
 (\mathbf{q}_5\times \mathbf{A}(\mathbf{q}_5))_c \tilde{A}_d(-\mathbf{q}_1-\mathbf{q}_2-\mathbf{q}_5)&.
\end{split}
\end{align}
For notational clarity, let us define the new variables:
\begin{align}
    \mathbf{p}=\mathbf{q}_1, && \mathbf{p}'=\mathbf{q}_5, && \mathbf{q} = -\mathbf{q}_1 - \mathbf{q}_2.
\end{align}

Making use of these variables, (\ref{eq:InteractionDerivation1}) becomes:
\begin{align}
    \label{eq:InteractionDerivation2}
- \int \dbar \mathbf{p} \dbar \mathbf{p}' \dbar \mathbf{q} \varepsilon_{jab} \varepsilon_{kcd}\gamma(\mathbf{q}) \sigma_{jk}(\mathbf{q}) (\mathbf{p} \times \mathbf{A}(\mathbf{p}))_a \tilde{A}_b(-\mathbf{p}-\mathbf{q})  (\mathbf{p}'\times \mathbf{A}(\mathbf{p}'))_c \tilde{A}_d(\mathbf{q}-\mathbf{p}').
\end{align}
%
%
More compactly, we can express
\begin{align}
\begin{split}
    S_{\text{int}} =-\frac{1}{2} \int \dbar \mathbf{p} \dbar \mathbf{p}' \dbar \mathbf{q} A_\alpha (\mathbf{p}) \tilde{A}_\beta (-\mathbf{p}-\mathbf{q}) M_{\alpha\beta\gamma\delta}^{kl}(\mathbf{q}) p_k p_l' A_\gamma (\mathbf{p}')\tilde{A}_\delta (\mathbf{q}-\mathbf{p}'),
\end{split}
\end{align}
where
\begin{align}
    M_{\alpha\beta\gamma\delta}^{kl}(\mathbf{q}) = \big( \delta_{k \beta} \delta_{l \delta} \sigma_{\alpha \gamma}(\mathbf{q}) - \delta_{\alpha \beta} \delta_{l\delta} \sigma_{k\gamma}(\mathbf{q})-\delta_{k\beta}\delta_{\gamma \delta} \sigma_{\alpha l} (\mathbf{q}) + \delta_{\alpha \beta} \delta_{\gamma \delta} \sigma_{k l} (\mathbf{q})  \big) \gamma(\mathbf{q}).
\end{align}

We may follow exactly the same procedure to Fourier transform in time. 
Reintroducing the time dependence, we find the interaction action to be of the form
\begin{align}
\begin{split}
    S_{\text{int}} =-\frac{1}{2} \int \dbar \mathbf{p} \dbar \mathbf{p}' \dbar \mathbf{q} \dbar \omega \dbar \omega' \dbar \Omega & \ A_\alpha (\mathbf{p},\omega) \tilde{A}_\beta (-\mathbf{p}-\mathbf{q},-\omega-\Omega) \\
    & M_{\alpha\beta\gamma\delta}^{kl}(\mathbf{q},\Omega) p_k p_l' A_\gamma (\mathbf{p}',\omega')\tilde{A}_\delta (\mathbf{q}-\mathbf{p}',\Omega - \omega'),
\end{split}
\end{align}
where $\omega,\omega'$ and $\Omega$ are Fourier conjugate frequencies and the $\Omega$ dependence in $M_{\alpha\beta\gamma\delta}^{kl}(\mathbf{q},\Omega)$ is entirely encapsulated by  $\gamma(\mathbf{q},\Omega)$.

\section{Gauge fixing term}\label{app:GaugeCheck}

We want to ensure that the addition of the gauge fixing term (\ref{eq:GaugeFix}) to the free decay action does not affect any of the magnetic field observables. For notational clarity, we keep time dependence implicit throughout this appendix.
Let us consider once again the expectation value of an observable $O[\mathbf{A}]$.
Computing the integral over $\mathbf{A}$ and $\tilde{\mathbf{A}}$ is relatively easy for $S_0[\mathbf{A},\tilde{\mathbf{A}}]$, since its form is Gaussian. This is not true for the interaction term. In order to make progress, we may expand $e^{-S_{\text{int}}}$ as a Taylor series, so the observable $\langle \mathbf{A} (\mathbf{q}) \rangle_u$ may be written to first order as
\begin{align}
   \langle A (\mathbf{q}) \rangle_u = \int \mathcal{D} \mathbf{A} \mathcal{D}\tilde{\mathbf{A}} \ A (\mathbf{q}) e^{-S_0[\mathbf{A},\tilde{\mathbf{A}}]}\bigg[1- S_{\text{int}}[\mathbf{A},\tilde{\mathbf{A}},\mathbf{C}] \bigg]+ O(S_{\text{int}}^2).
\end{align}
A general observable of $\mathbf{B}$ may be computed via
\begin{align}
    \langle \mathbf{B} (\mathbf{q}_1) \dots \mathbf{B} (\mathbf{q}_n)\rangle_u =  \int \mathcal{D} \mathbf{A} \mathcal{D} \tilde{\mathbf{A}} \big( \mathbf{q}_1 \times \mathbf{A}(\mathbf{q}_1)\dots \mathbf{q}_n \times \mathbf{A}(\mathbf{q}_n) \big) e^{-S_0[\tilde{\mathbf{A}},\mathbf{A}]}\bigg[1- S_{\text{int}}[\mathbf{A},\tilde{\mathbf{A}}] \bigg]
\end{align}

Given that the probability distribution for $\mathbf{u}$ is Gaussian, we may apply Wick's theorem, which states that
\begin{align}
    \Bigg\langle \prod_{i=1}^\ell \mathbf{A}_i(\mathbf{q}_i) \Bigg\rangle_u = 
    \begin{dcases}
    0 & \ell \text{ odd},\\
    \text{sum over pairwise contractions} & \ell \text{  even}.
    \end{dcases}
\end{align}
Using Wick's theorem we may consider purely pairwise correlations, since we are able to express any observable in terms of them. This means that we need only consider the following three pairwise correlations to ensure that the gauge fixing term has no effect on the correlations of physical observables (since the variables commute):
\begin{align}
    \langle \tilde{\mathbf{A}}(\mathbf{q}) \tilde{\mathbf{A}}(\mathbf{q}') \rangle_u, && \langle \tilde{\mathbf{A}}(\mathbf{q}) \mathbf{A} (\mathbf{q}')\rangle_u, &&  \langle \mathbf{A}(\mathbf{q}) \mathbf{A}(\mathbf{q}') \rangle_u.
\end{align}
Writing the action $S_0[\mathbf{A},\tilde{\mathbf{A}}]$ in matrix form, 
\begin{align}
\label{eq:S0Matrix}
 S_0[\mathbf{A},\tilde{\mathbf{A}}] =\frac{1}{2} \mathbf{a}_\alpha \Gamma^{-1}_{\alpha \beta} \mathbf{a}_\beta = \frac{1}{2}
\begin{pmatrix}
  A_\alpha &
   \tilde{A}_\alpha
\end{pmatrix}
    \begin{pmatrix}
     \frac{i}{g}q_\alpha q_\beta & -i G_0^{-1} \delta_{\alpha \beta}\\
    -i G_0^{-1} \delta_{\alpha \beta} & 0
    \end{pmatrix}
    \begin{pmatrix}
  A_\beta\\
   \tilde{A}_\beta
\end{pmatrix},
\end{align}
where $G_0^{-1}(k,\omega) = i\omega + \eta k^2$,
allows us to read off these pairwise correlations, given by 
\begin{align}
\label{eq:AcorrsS0}
    \langle A_\alpha A_\beta \rangle = 0, &&
    \langle A_\alpha \tilde{A}_\beta \rangle = i G_0 \delta_{\alpha \beta},&&     \langle \tilde{A}_\alpha \tilde{A}_\beta \rangle = \frac{i}{g}G_0q_\alpha q_\beta G_0.
\end{align}
Let us consider the lowest order case that contains the interaction term for:
\begin{align}
\begin{split}
    \langle \mathbf{B}_\alpha(\mathbf{k}) \mathbf{B}_\beta(\mathbf{k}')\rangle = &- \int \mathcal{D} \mathbf{A} \mathcal{D} \tilde{\mathbf{A}}  \big( \mathbf{k}\times \mathbf{A})_\alpha(\mathbf{k}' \times \mathbf{A})_\beta e^{-S_0[\tilde{\mathbf{A}},\mathbf{A}]}\\
    &\bigg[ 1 +\frac{1}{2} \int \dbar \mathbf{p} \dbar \mathbf{p}' \dbar \mathbf{q} A_\alpha (\mathbf{p}) \tilde{A}_\beta (-\mathbf{p}-\mathbf{q}) M_{\alpha\beta\gamma\delta}^{kl}(\mathbf{q}) p_k p_l' A_\gamma (\mathbf{p}')\tilde{A}_\delta (\mathbf{q}-\mathbf{p}')\bigg].
\end{split}
\end{align}
According to Wick's theorem we can split the integral into a sum of pairwise interactions between the $\mathbf{A}$ and $\tilde{\mathbf{A}}$ terms. According to (\ref{eq:AcorrsS0}), the only interaction that appears to depend on the choice of gauge is $\langle \tilde{A}_\alpha \tilde{A}_\beta \rangle$, so we need only consider this combination. We find that 
\begin{align}
   M_{\alpha\beta\gamma\delta}^{kl}(\mathbf{q})\langle \tilde{A}_\beta (-\mathbf{p}-\mathbf{q}) \tilde{A}_\delta (\mathbf{q}-\mathbf{p}') \rangle = -\frac{i q^2}{g} G_0^2 \sigma_{\alpha\gamma} (\mathbf{q}).
\end{align}
Note that $q_j \varepsilon_{jab} q_b= 0$ by symmetry.
Since there was only one pair of $\tilde{\mathbf{A}}$s, this operator can now only act on $\langle \mathbf{A} \mathbf{A}\rangle$ correlations. Therefore, we have
\begin{align}
   \frac{i q^2}{g} G_0^2 \sigma_{\alpha\gamma} (\mathbf{q}) \langle A_\alpha (\mathbf{p}) A_\gamma (\mathbf{p}') \rangle = 0,
\end{align}
which vanishes thanks to the matrix element in $\Gamma_{\alpha \beta}$. We can extend this logic to more complex cases by noting that there will always be more pairs of $A$s than $\tilde{A}$s, so if any $\tilde{A}$s are paired up with each other we will inevitably end up with a $\langle A_\alpha A_\beta \rangle$ that causes the term to vanish. The only terms that do not vanish involve correlations of $A$ with $\tilde{A}$, so there is no dependence of the physical observables on $g$, and by extension on the choice of gauge. This is the Ward-Takahashi identity in quantum electrodynamics\cite{PeskinSchroder}.

\section{Magnetic field Green function for all loop orders} \label{app:AllLoopOrdDiffCalc}

To calculate the all orders loop contribution to the propagator, we must assume that the loops dominate over the other 1PI diagrams, and that this is true at all orders (i.e. that a string of $n$ loops dominates over any other $n$-vertex diagram).
\begin{figure}
    \centering
    \includegraphics[width=\textwidth]{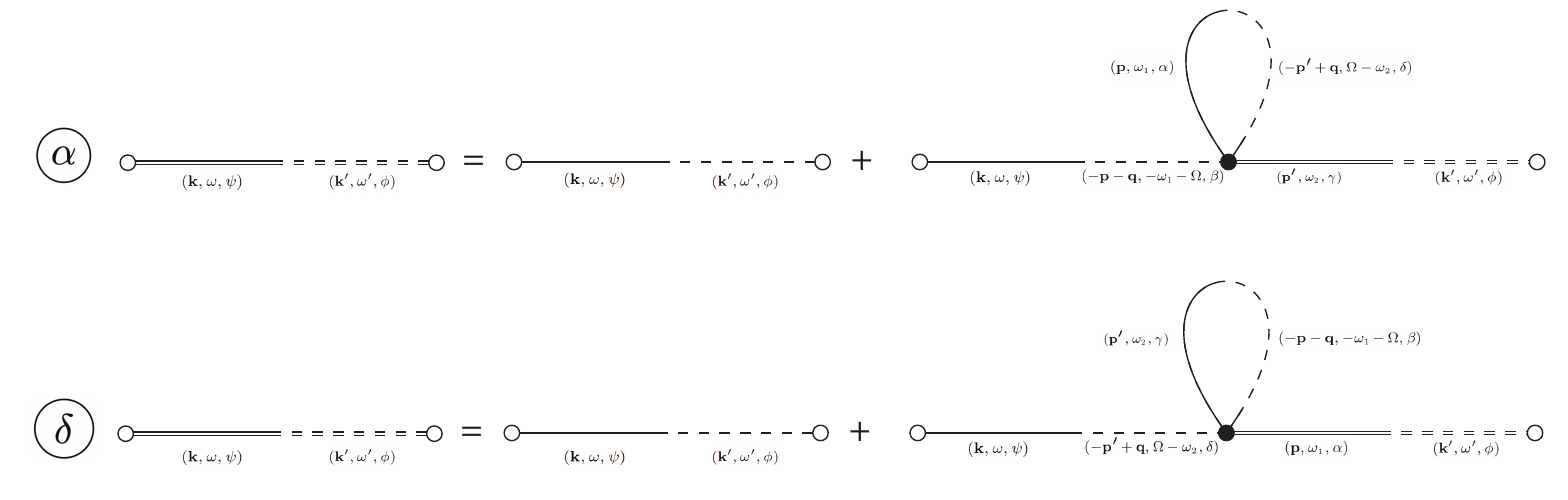}
    \caption{Diagrammatic illustration of the expansion for the quantity $\langle \mathbf{A}(\mathbf{k}) \tilde{\mathbf{A}}(\mathbf{k}')\rangle$. Note that the joined pairs of double lines do not contribute a zero order propagator $G_0$, but rather contribute the exact propagator $\Gamma$. Only two diagrams,\protect\circled{$\alpha$} and \protect\circled{$\delta$}, contribute, as in section 4\ref{sec:DiagsIntCorr}.
    }
    \label{fig:AllOrdersLoop}
\end{figure}
Taking $\boldsymbol{\Xi}$ to first order so that it only includes loop contributions, fig. \ref{fig:1PIAllOrders} tells us that the Green function ($i\Gamma_{\psi \phi}$) for $\mathbf{A}$, may be written as
\begin{flalign}
\label{eq:GammaPsiPhi}
\begin{split}
    i \Gamma_{\psi \phi} (\mathbf{k},\mathbf{k}',\omega,\omega') = &i G_0(\mathbf{k},\omega)\delta(\mathbf{k}+\mathbf{k}')\delta(\omega+\omega') \delta_{\psi \phi}\\
    &- \frac{i}{2}\int \dbar \mathbf{q} \dbar \mathbf{p} \dbar \mathbf{p}' \dbar \omega_1 \dbar \omega_2 \dbar \Omega \
     \delta(\mathbf{k}-\mathbf{p}-\mathbf{q})\delta(\mathbf{q}-\mathbf{p}'+\mathbf{p}) \delta(\omega-\omega_1-\Omega)\\
     &\hspace{2.2cm}\delta(\Omega-\omega_2+\omega_1)
    G_0(\mathbf{k},\omega) G_0(\mathbf{p},\omega_1) \Gamma_{\gamma \phi} (\mathbf{p}',\mathbf{k}',\omega_2,\omega')\\
    &\hspace{6.1cm} \delta_{\psi \beta} \delta_{\alpha \delta} p_k p_l' M_{\alpha \beta \gamma \delta}^{kl}(\mathbf{q},\Omega)  \\
    &-\frac{i}{2}\int \dbar \mathbf{q} \dbar \mathbf{p} \dbar \mathbf{p}' \dbar \omega_1 \dbar \omega_2 \dbar \Omega \ \delta(\mathbf{k}-\mathbf{p}'+\mathbf{q})\delta(\mathbf{q}-\mathbf{p}'+\mathbf{p}) \delta(\omega-\omega_2+\Omega)\\
    & \hspace{2.5cm}\delta(\Omega-\omega_2+\omega) G_0(\mathbf{k},\omega) G_0(\mathbf{p}',\omega_2) \Gamma_{\alpha \phi} (\mathbf{p},\mathbf{k}',\omega_1,\omega') \\
    & \hspace{6.2cm} \delta_{\psi \delta} \delta_{\beta \gamma} p_k p_l' M_{\alpha \beta \gamma \delta}^{kl}(\mathbf{q},\Omega) ,
\end{split}
\end{flalign}
where the terms on the RHS are the zero-order, \circled{$\alpha$} and \circled{$\delta$} contributions respectively. 
The calculation goes through in much the same way as eqs. (\ref{eq:AlphaContribution}) - (\ref{eq:DeltaContribution}), yielding
\begin{flalign}
    \circled{$\alpha$}
    =& - \frac{i}{2}\int \dbar \mathbf{q} \dbar \Omega \ G_0(\mathbf{k},\omega) G_0(\mathbf{k}-\mathbf{q},\omega-\Omega)
    k_l(\mathbf{k}-\mathbf{q})_k M_{\alpha \psi \gamma \alpha}^{kl}(\mathbf{q},\Omega) \Gamma_{\gamma \phi} (\mathbf{k},\mathbf{k}',\omega,\omega').\\
    \circled{$\delta$} = & -\frac{i}{2} \int \dbar \mathbf{q} \dbar \Omega \ G_0(\mathbf{k},\omega) G_0(\mathbf{k}+\mathbf{q},\omega+\Omega)  \bigg(k^2 - \frac{1}{q^2}(\mathbf{k}\cdot\mathbf{q})^2 \bigg) \gamma(\mathbf{q},\Omega) \Gamma_{\psi\phi} (\mathbf{k},\mathbf{k}',\omega,\omega').
\end{flalign}

Equation (\ref{eq:GammaPsiPhi}) may then be written as
\begin{align}
\begin{split}
    & G_0(\mathbf{k},\omega)\delta(\mathbf{k}+\mathbf{k}') \delta(\omega+\omega')\delta_{\psi \phi} =\Gamma_{\psi \phi} (\mathbf{k},\mathbf{k}',\omega,\omega')\\
    &\Bigg[1+ \int \dbar \mathbf{q} \dbar \Omega \ \frac{\gamma(\mathbf{q},\Omega)}{2} G_0(\mathbf{k},\omega) \left(G_0 (\mathbf{k}-\mathbf{q},\omega-\Omega)+G_0 (\mathbf{k}+\mathbf{q},\omega+\Omega)\right) \bigg(k^2 - \frac{(\mathbf{k}\cdot\mathbf{q})^2}{q^2}\bigg) \Bigg],
\end{split}
\end{align}
where the $\Gamma_{\psi \phi}$ terms have been collected on the LHS. It follows that
\begin{align}
\label{eq:AsGreensFunctAllOrd}
     \Gamma_{\psi \phi} (\mathbf{k},\mathbf{k}',\omega,\omega')=\frac{G_0(\mathbf{k},\omega)\delta(\mathbf{k}+\mathbf{k}')\delta(\omega+\omega') \delta_{\psi \phi}}{1+G_0(\mathbf{k},\omega)\Xi(\mathbf{k},\omega)},
\end{align}
where
\begin{align}
    \Xi (\mathbf{k},\omega) = \frac{1}{2} \int \dbar \mathbf{q} \dbar \Omega \ \gamma(\mathbf{q},\Omega) \big[G_0 (\mathbf{k}-\mathbf{q}, \omega-\Omega)+G_0 (\mathbf{k}+\mathbf{q},\omega+\Omega)\big] \bigg(k^2 - \frac{1}{q^2}(\mathbf{k}\cdot\mathbf{q})^2 \bigg).
\end{align}
Equation (\ref{eq:AsGreensFunctAllOrd}) enables us to compute the magnetic vector potential's Green function to all orders in the loop integral. To obtain the magnetic field's Green function we make use of (\ref{eq:GAtoGB}) again, noting that this time it accounts for the contribution of the loop diagrams to all orders
\begin{align}
\begin{split}
\mathcal{G}_{\alpha \beta}(\mathbf{k},\mathbf{k}',\omega,\omega') =&\frac{(2\pi)^3}{k^2} \delta(\mathbf{k}+\mathbf{k}')\delta(\omega+\omega') \varepsilon_{\beta n \phi} \varepsilon_{\alpha l \psi} k_n k_l \bigg(\delta_{\psi \phi} \frac{G_0(\mathbf{k},\omega)}{1+G_0(\mathbf{k},\omega)\Xi(\mathbf{k},\omega)} \bigg)\\
=&\frac{(2\pi)^3\delta(\mathbf{k}+\mathbf{k}') \delta(\omega+\omega') \sigma_{\alpha\beta}(\mathbf{k}) \delta_{\psi \phi}}{G_0(\mathbf{k},\omega)^{-1}+\Xi(\mathbf{k},\omega)}.
\end{split}
\end{align}
 
\bibliography{references}
\end{document}